\documentclass[pra,twocolumn,showpacs,preprintnumbers,amsmath,amssymb,floatfix]{revtex4-1}

\usepackage{graphicx}
\usepackage{dcolumn}
\usepackage{bm}
\usepackage{color}
\usepackage{mathrsfs}

\newcommand{\bra}[1]{\langle #1|}
\newcommand{\ket}[1]{|#1\rangle}

\newcommand{\affA}{Zentrum f\"ur Optische Quantentechnologien, Fachbereich Physik, Luruper Chaussee 149, D-22761 Hamburg, Germany}

\usepackage{ulem}
\usepackage{physics}
\usepackage{multirow}

\usepackage{makecell}

\begin{document}


\title{Quantum-limited thermometry of a Fermi gas with a charged spin particle}


\author{Lorenzo Oghittu and Antonio Negretti}
\affiliation{\affA}

\date{\today}


\begin{abstract}
We investigate the sensitivity of an ion sensor in determining the temperature of an atomic Fermi gas. Our study extends to charged impurities the proposal by M. T. Mitchison \textit{et al.} Phys. Rev. Lett. {\bf 125}, 080402 (2020), where atomic neutral impurities were used as an \textit{in situ} thermometer of the quantum gas. We find that the long-range character of the atom-ion interaction enhances the thermometer's sensitivity for certain system parameters. In addition, we investigate the impact of the ion quantum motional state on the sensitivity by assuming that it is confined in a harmonic trap. We observe that the temperature sensitivity of the ion is noticeably influenced by its spatial extension, making the latter a versatile tool to be manipulated for improving the thermometer performance. We finally discuss our findings in the context of current experimental atom-ion mixtures. 
\end{abstract}

\maketitle


\section{Introduction}
Quantum impurities in a bath such as that provided by ultracold atomic gases is currently a very active field of research owed to the advances and prospects offered by atomic physics laboratories. Indeed, these systems enable with their degree of controllability to investigate various out-of-equilibrium phenomena such as Anderson orthogonality catastrophe~\cite{anderson1967a,Demler_PRX12} and the formation of polaronic states~\cite{Devreese2009,Alexandrov2010}, to mention a few. The former is fundamental for understanding phenomena in the solid-state realm like the Kondo effect~\cite{anderson1970} and transport of heavy impurities in a Fermi liquid~\cite{Rosch1999}, while mediated interactions are important for pairing formation~\cite{2003,Alexandrov2011,scalapino2018}. In particular, in recent years the study of Fermi polarons with neutral impurities in bulk systems has been quite vigorous both experimentally~\cite{Koehl_NAT12,Cetina_SC16,Roati_PRL17,Zaccanti_PRL20} and theoretically~\cite{Massignan_2014,Schmidt_2018,atoms9010018}. Albeit not yet mature as the neutral counterpart, an important experimental effort has been undertaken to immerse charged impurities such as ions in quantum gases~\cite{HarterCP14,CoteAAMOP16,TomzaRMP}. Sympathetic cooling of ions in a Fermi gas with calcium ions has been experimentally investigated~\cite{Giapponesi_PRL18}, which culminated with the approach of the $s$-wave regime of atom-ion collisions~\cite{Gerritsma_NAT20, Schaetz_NAT21}. In addition to the aforementioned out-of-equilibrium phenomena, ions in a 1D fermionic bath have been proposed to study induced interactions in a Li-Yb$^+$ mixture~\cite{Michelsen2019} (see also Ref.~\cite{Mistakidis_2019} for neutral ytterbium impurities), Peierls instability~\cite{Bissbort2013}, and bipolaron states with low effective mass~\cite{JachymskiPRR2020}. Such interesting many-body quantum physics, however, requires a Fermi gas to reach very low temperatures, i.e., on the order of a few percentage of the Fermi temperature. It is therefore crucial to devise experimental schemes to attain those temperatures as well as to determine accurately its uncertainty. 

Recently, M.~T. Mitchison \textit{et al.}~\cite{Goold_PRL20} have proposed an interferometric method to probe locally the temperature of a Fermi gas by means of a neutral spin impurity. The method relies on the fact that the Pauli exclusion principle slows the decay of coherence of the spin impurity allowing for enhanced signal-to-noise ratios. Therefore, by tracking the dynamics of the spin impurity and, importantly, without requiring its thermalisation, it is possible to estimate the temperature with high accuracy.
Here, we extent the method to an ionic impurity probe, whose impurity-gas interaction is longer-ranged compared to the neutral case. We find that the $r^{-4}$ tail of the  atom-ion polarization potential has a profound impact on the thermometer sensitivity, quantified by the quantum signal-to-noise ratio $\mathcal{Q}$, and that for some system parameters it reaches a larger performance than a neutral probe. We investigate $\mathcal{Q}$ in reliance of the density of the gas, the number of two-body bound states, and finite-width of the ion spatial density, i.e., confinement. The latter is shown to enable to obtain higher signal-to-noise ratios and at shorter times. 
Let us also note that ions have been already used as probes of the density profile of a condensate~\cite{SchmidPRL10,ZipkesPRL10} as well as to trace molecule gases~\cite{Gerritsma_arXiv21} and proposed for measuring density-density correlations~\cite{KollathPRA07} and the local single-particle energy distribution of a degenerate Fermi gas~\cite{SherkunovPRA09}.

The paper is organized as follows: in Sec.~\ref{sec:System} we describe our system, while in Sec.~\ref{sec:Thermometry} the Cramer-Rao bound together with the interferometric protocol for sensing the gas temperature is summarised. This summary, which is based on the ideas of Ref.~\cite{Goold_PRL20} together with a few remarks from our side, is provided for the sake of completeness. The results are exposed and discussed in Sec.~\ref{sec:Results}, whereas the experimental applicability of the extended method to the ionic probe is discussed in Sec.~\ref{sec:Discussion}. Finally, in Sec.~\ref{sec:Conclusions} we recapitulate our findings and provide an outlook for future work.


\section{System}
\label{sec:System}
\begin{figure}
    \centering
    \includegraphics[width=.5\textwidth]{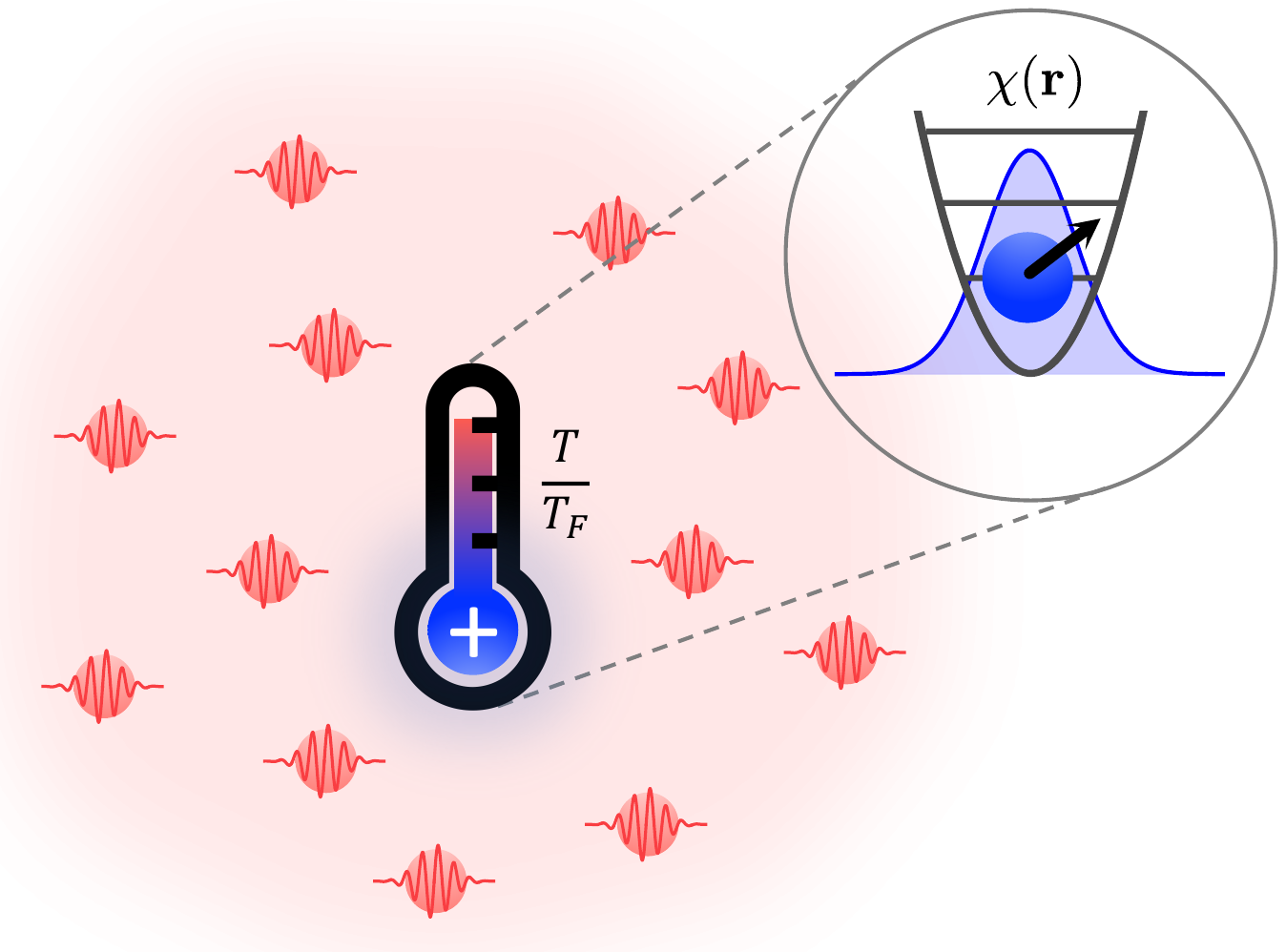}
    \caption{(Color Online). Schematic of the system: a positively charged ion, immersed in a Fermi gas (red wavy particles), acts as a thermometer for measuring the gas temperature ratio $T/T_\mathrm{F}$. As it is shown in the inset, such an ion is harmonically trapped with Gaussian spatial density $\vert\chi(\mathbf{r})\vert^2$ (blue shadowed area), i.e., it is prepared in the ground state, and has a spin degree of freedom denoted pictorially by an arrow.}
    \label{fig:system}
\end{figure}
\begin{figure}
    \centering
    \includegraphics[width=.4\textwidth]{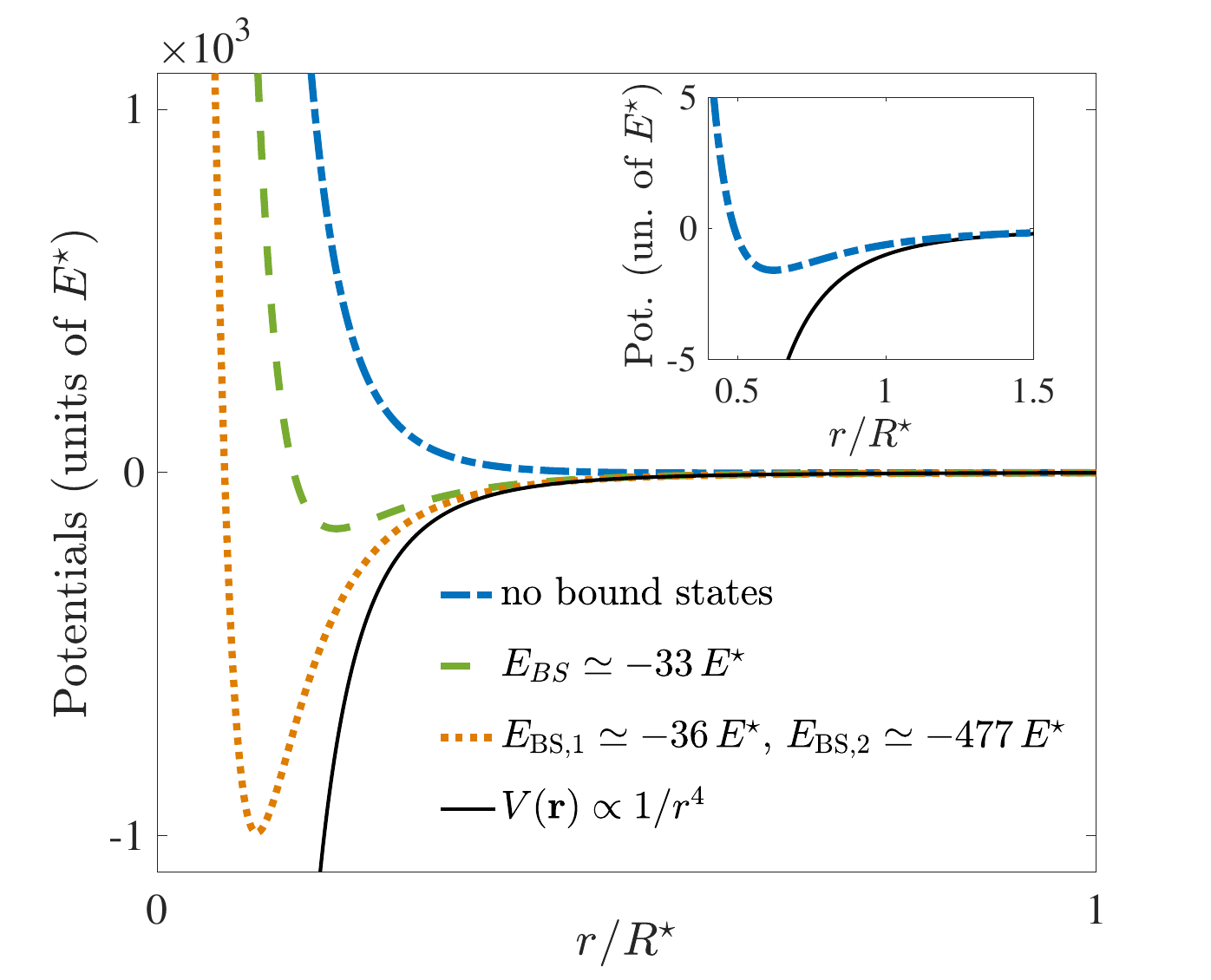}
    \caption{(Color online) Regularized atom-ion polarization potentials $V_\mathrm{reg}(r)$ corresponding to $a\simeq-R^\star$ with three different bound state properties: no bound states in blue dot-dashed, one bound state in green dashed and two bound states in orange dotted (see legend for bound state energies). Two-body polarization potential~(\ref{eq:singular_potential}) in thin black solid. The inset illustrates the $r^{-4}$-tail of the potential without bound states.}
    \label{fig:potentials}
\end{figure}

We consider the system displayed pictorially in Fig.~\ref{fig:system}. 
A (trapped) ion (or more), whose wave function is denoted by $\chi(\mathbf{r})$, has two internal states $\ket{0}$ and $\ket{1}$. The ion is immersed in a homogeneous spin-polarized non-interacting Fermi gas (henceforth also referred to as bath) of mean density $\Bar{n}$. The gas density defines the following bath characteristics: the Fermi wave vector $k_\mathrm{F}=(6\pi\Bar{n})^{1/3}$, the energy (or temperature) $E_\mathrm{F}=\hbar^2k_\mathrm{F}^2/2m = k_\mathrm{B} T_\mathrm{F}$, and the time $\tau_\mathrm{F}=\hbar/E_\mathrm{F}$. Here, $k_\mathrm{B}$ is the Boltzmann constant and $m$ the mass of the atom. 
In the case of a neutral impurity probe~\cite{Goold_PRL20}, one can assume that the impurity internal state $\ket{0}$ does not interact with the bath, while the state $\ket{1}$ interacts via a short-range impurity-bath pseudopotential. This is legitimate, because one can tune the impurity-bath interaction such that the scattering length vanishes, thereby resulting effectively in a vanished potential~\footnote{This is permitted because of the separation of length scales involved in the system, that is, the range of the impurity-bath interaction is much smaller than any other length scale.}. In the present setting, however, the electric field of the ion polarizes the neighbourhood independently of the atom-ion scattering length such that the $r^{-4}$-tail of the spin-independent polarization potential cannot be set to zero in our model. In other words, that tail is always present, unless the gas density is so low that the interaction can be replaced by a pseudopotential, thus falling back again into the neutral probe scenario. Controlling the atom-ion scattering length and therefore the ion internal state means that we manipulate the short-range part of the potential only (e.g., the number of bound states). Given this, both ion internal states interact (asymptotically) with the bath via the two-body polarization potential
\begin{equation}
    V(\mathbf{r})=-\frac{C_4}{r^4}
    \label{eq:singular_potential}
\end{equation}
with $C_4=\alpha e^2/8\pi\epsilon_0$ (in SI units). Here, $\alpha$ is the (static) polarizability of the atom, $e$ is the elementary electronic charge (i.e., only singly ionized atoms) and $\epsilon_0$ the vacuum permittivity. We note that recently Feshbach resonances in hybrid atom-ion systems have been observed~\cite{Schaetz_NAT21}, thus providing the perspective to control (magnetically) the interspecies interactions.\\
The potential~(\ref{eq:singular_potential}) introduces two more length and energy scales: $R^\star=(2\mu_\mathrm{red} C_4/\hbar^2)^{1/2}$ and $E^\star=\hbar^2/[2\mu_\mathrm{red}(R^\star)^2]$ with $\mu_\mathrm{red} = m M /(m + M)$ being the reduced atom-ion mass and $M$ is the ion mass. 
Due to the singularity of Eq.~\eqref{eq:singular_potential} at short range and for the sake of a simpler analytical treatment, we use the regularization~\cite{Idziaszek_PRA15, Oghittu_PRA21}
\begin{equation}
    V_\mathrm{reg}^{(s)}(\mathbf{r})=-C_4\frac{r^2-c^2}{r^2+c^2}\frac{1}{(r^2+b^2)^2}.
    \label{eq:potential}
\end{equation}
An example of the spatial dependence of this potential is shown in Fig.~\ref{fig:potentials}. By tuning the parameters $b$ and $c$ one can control the sign and strength of the three-dimensional $s$-wave atom-ion scattering length $a$ as well as the number of bound states, which can be attained by preparing the ion in specific electronic configurations, i.e., spin state $\ket{s}$, and by controlling external magnetic fields. Towards this end, we employ the strategy already used in Ref.~\cite{Oghittu_PRA21}, namely we impose that the scattering length equals in magnitude the scattering amplitude of the regularized potential in Born approximation and that the potential supports a fixed number of bound states (one or two). For the case without bound states, we just seek for parameters that generate a potential without bound states, but that yield a certain value of the scattering length.

For the purpose of thermometry, we are interested mainly in the dynamics of the ion's internal degrees of freedom, since these two states are utilized to read out the gas temperature and its uncertainty, while the motional degrees of freedom affect the ion-gas interaction, as we shall see later. Hence, the ionic probe internal state Hamiltonian reads as~\footnote{We note that we do not include the motional part of the ion, i.e., its (spin-independent) Hamiltonian in the trap, as we consider it as a static impurity, namely a mere scattering centre for the bath. Even in the non static approximation, however, that Hamiltonian, being spin-independent, would simply add a global phase that can be removed by moving the system description in the interaction picture. We can thus safely ignore it.}
\begin{equation}
    \hat{H}_\mathrm{P}=E_0\ket{0}\bra{0}+E_1\ket{1}\bra{1}.
    \label{eq:H_P}
\end{equation}
On the other hand, the many-body Hamiltonian of the bath is given by
\begin{equation}
    \hat{H}_\mathrm{B}=\int_{\mathbb{R}^3}d\mathbf{r}\,\hat\Psi^\dag(\mathbf{r})\bigg[-\frac{\hbar^2}{2m}\nabla^2 + V_\mathrm{ext}(\mathbf{r})\bigg]\hat\Psi(\mathbf{r}),
    \label{eq:H_B}
\end{equation}
where $V_\mathrm{ext}(\mathbf{r})$ denotes the external trap potential, $\hat\Psi$ is the fermionic antisymmetric field operator satisfying the anti-commutation rule $\{\hat\Psi(\mathbf{r}),\hat\Psi^\dag(\mathbf{r}')\}=\delta(\mathbf{r}-\mathbf{r}')$.  
The (many-body) interaction between the ion and the quantum gas is described by the Hamiltonian
\begin{equation}
    \hat{H}_\mathrm{I}=\sum_{s = 0,1}\int_{\mathbb{R}^3}d\mathbf{r}\,\hat\Psi^\dag(\mathbf{r})V^{(s)}_\mathrm{reg}(\mathbf{r})\hat\Psi(\mathbf{r})\otimes\ket{s}\bra{s},
    \label{eq:H_I}
\end{equation}
where $V^{(s)}_\mathrm{reg}$ relies on the ion internal state via the short-range physics we discussed earlier, namely by tuning the atom-ion scattering length with different choices of the parameters $b,\,c$. Moreover, we note that $[\hat{H}_\mathrm{P},\hat{H}_\mathrm{I}]=0$. 

\begin{figure}
    \centering
    \includegraphics[width=0.45\textwidth]{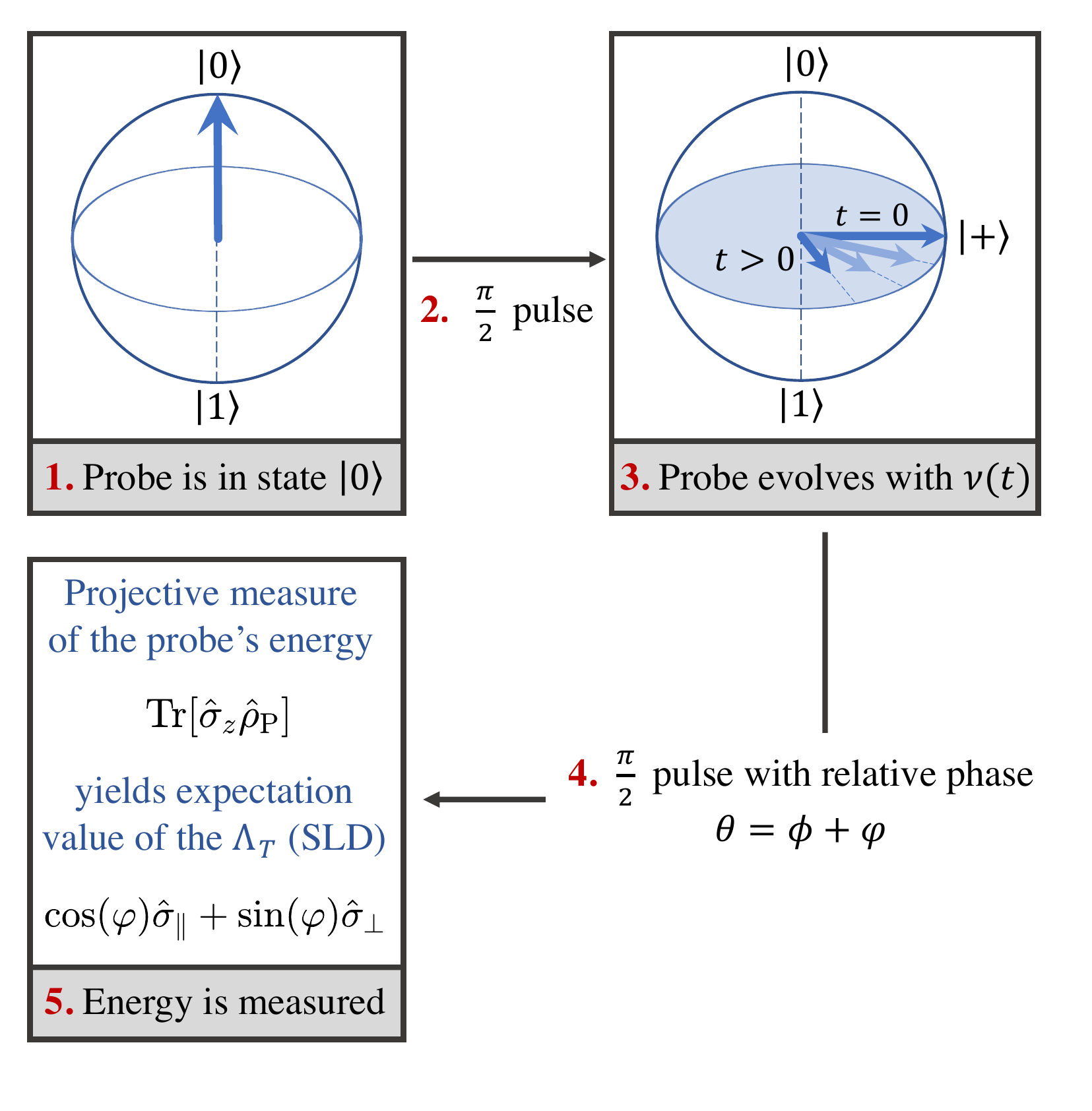}
    \caption{(Color Online) Schematic of the interferometric protocol with its five steps that allow to determine the expectation value of the SLD. The SLD is the estimator of the temperature that maximizes the quantum Fisher information.}
    \label{fig:protocol}
\end{figure}

The ion spin dynamics can be described by a single time-dependent function, as we show in some detail in appendix~\ref{sec:Calculations}. Specifically, the ion Bloch vector $\mathbf{v}$, whose density matrix is $\hat{\rho}_\mathrm{P}=\frac{1}{2}(\mathbb{I}+\mathbf{v}\cdot\hat{\boldsymbol{\sigma}})$, is given by~\footnote{Here, we implicitly assume that the ion internal state lies in the equatorial plane as shown in Fig.~\ref{fig:protocol} (see protocol in Sec.~\ref{sec:Thermometry}). Let us also note that the populations of the eigenstates of $\hat\sigma_x$ are conserved as a consequence of $[\hat{H}_\mathrm{P},\hat{H}_\mathrm{I}]=0$.}
\begin{equation}
\label{eq:blochvec}
\mathbf{v}(t)\equiv (v_x, v_y, v_z) =(\mathrm{Re}[\nu(t)],\mathrm{Im}[\nu(t)],0), 
\end{equation}
where the so-called time-dependent decoherence function is defined as
\begin{equation}
    \nu(t)=\mathrm{Tr}_\mathrm{B}\Big[e^{i\hat{H}_1 t/\hbar}e^{-i\hat{H}_0 t/\hbar}\hat{\rho}_\mathrm{B}(T)\Big]
    \label{eq:decoherence}
\end{equation}
with $\hat{H}_s=\bra{s}\hat{H}_\mathrm{B}+\hat{H}_\mathrm{I}\ket{s}$, $s=0,1$, $\hat{\rho}_\mathrm{B}(T)$ the thermal state at temperature $T$ of the bath, and 
$\hat{\boldsymbol{\sigma}}=(\hat{\sigma}_x,\hat{\sigma}_y,\hat{\sigma}_z)$ the vector of Pauli matrices. We note that since the $z$-component of the Bloch vector is zero [see Eq.~\eqref{eq:blochvec}], the populations of the states $\ket{0}$ and $\ket{1}$ are conserved in time and equal to 1/2, while the coherences evolve accordingly to Eq.~\eqref{eq:decoherence} (pure dephasing). Notwithstanding, when $\pi/2$-pulses are applied to ion internal state prior and after the impurity and the bath interact, as we shall discuss in the interferometric protocol of Sec.~\ref{sec:Thermometry}, $v_z(t)$ relies on $\nu(t)$. 

By means of the well-known Levitov formula~\cite{Levitov_PRL05,Muzykantskii_PRB05}, the decoherence function can be computed exactly via the formula~\footnote{We note that the Levitov formula can be applied to quadratic and time-independent Hamiltonians only.}
\begin{equation}
    \nu(t)=\mathrm{det}\Big[1-\hat{n}+\hat{n}e^{i\hat{h}_0t/\hbar}e^{-i\hat{h}_1t/\hbar}\Big],
    \label{eq:simplified_nu}
\end{equation}
where $\hat{n}=(e^{\beta(\hat{h}_\mathrm{B}-\mu)}+1)^{-1}$ is the Fermi distribution, $\mu$ denotes the chemical potential, $\beta = (k_\mathrm{B} T)^{-1}$, and 
\begin{equation}
    \begin{split}
        \hat{h}_\mathrm{B}&=-\frac{\hbar^2}{2m}\nabla^2+V_\mathrm{ext}(\mathbf{r}),\\[2ex]
        \hat{h}_\alpha&=\hat{h}_B+V^{(s)}_\mathrm{eff}(\mathbf{r}) \text{ with $s = 0,1$}.
    \end{split}
    \label{eq:h1}
\end{equation}
In our setting the Fermi gas is homogeneous and as a consequence $V_\mathrm{ext}=0$. The effective impurity-gas interaction is given by
\begin{equation}
\label{eq:convolution}
    V^{(s)}_\mathrm{eff}(\mathbf{r})=\int\mathrm{d}\mathbf{r}'V^{(s)}_\mathrm{reg}(\mathbf{r}-\mathbf{r}')|\chi(\mathbf{r}')|^2.
\end{equation}
Here, the reliance of $V_\mathrm{reg}^{(s)}$ on the ion internal state $\ket{s}$ highlights the fact that we can associate to it two scattering lengths for different pairs $(b, c)$.
As it can be seen, the ion motional state $\chi(\mathbf{r}')$ determines the shape of the effective interaction (see also Fig.~\ref{fig:effective_potentials} in Sec.~\ref{sec:Results_LowT}).
In the following sections we shall consider two different forms of the ionic probe's probability density $|\chi(\mathbf{r})|^2$. Firstly, we shall treat the ion as a static point particle and set $|\chi(\mathbf{r})|^2=\delta(\mathbf{r})$ or, in spherical coordinates
\begin{equation}
    |\chi(\mathbf{r})|^2=\frac{1}{4\pi r^2}\delta(r).
    \label{eq:chi_delta}
\end{equation}
Secondly, we shall replace the point-like distribution of Eq.~\eqref{eq:chi_delta} with a finite-width Gaussian, that is, $\delta(r)\mapsto\Delta_\sigma(r)=\exp[-r^2/(2\sigma^2)]/\sqrt{2\pi\sigma^2}$. In the limit 
$\sigma\to0$, we retrieve Eq.~(\ref{eq:chi_delta}) for which $V^{(\alpha)}_\mathrm{eff}(\mathbf{r})=V^{(\alpha)}_\mathrm{reg}(\mathbf{r})$. We make use of a Gaussian spatial density as it represents the ground state of an ion in a Paul trap~\cite{Blatt_RMP03} (i.e., secular approximation) as well as the one in an optical trap~\cite{Schaetz_NP10}.


\section{Thermometry with a spin impurity}
\label{sec:Thermometry}
In this section we recap the key aspects of the interferometric protocol proposed in Ref.~\cite{Goold_PRL20}. 
For more details, however, we refer to that paper. \\[1ex]
\textit{Quantum Cramer-Rao bound. $-$} In order to motivate the protocol discussed below, it is necessary to resort to the theory of quantum parameter estimation \cite{Paris_IJQI09,Apellaniz_IOP14,Wasak2016,Correa_IOP19}. Let us start by invoking the so-called quantum Cramer-Rao bound (QCRB). This provides a bound (from below) on the attainable uncertainty $\Delta T$ for the estimation of the temperature of the gas after $N$ independent realizations, i.e., measurements, according to $\Delta T^2\geq1/N\mathcal{F}_T\geq1/N\mathcal{F}_T^Q$, which is valid for any unbiased estimator~\cite{Caves_PRL94}. Note, however, that the first inequality, i.e. $\Delta T^2\geq1/(N\mathcal{F}_T)$, is reached for $N\gg 1$ within the maximum likelihood estimation procedure~\cite{refregier2012noise}. Here, $\mathcal{F}_T$ is the (classical) Fisher information~\cite{refregier2012noise}
\begin{equation}
    \mathcal{F}_T \equiv \mathcal{F}_T(\hat X) = - \sum_{s = \pm} p(x_s\vert T) \frac{\partial^2}{\partial T^2} \ln[p(x_s\vert T)]
    \label{eq:fisher}
\end{equation}
with $\hat X$ being any hermitian operator (i.e., observable), whereas $p(x_s\vert T)$ is the conditional probability to obtain from a measurement the outcome $x_s$, i.e., an eigenvalue of $\hat X$, given the gas temperature $T$. Let us underscore that the probability $p(x_s\vert T)$ results from measurements described by projections (or more generally by positive operator-valued measurement operators) on the corresponding eigenspaces of $\hat X$. Furthermore, since we consider a two-level system, not more than two outcomes, i.e., eigenvalues of $\hat X$, are possible. 
Given this, $\mathcal{F}_T$ is related to its quantum counterpart $\mathcal{F}_T^Q$ by the identity $\mathcal{F}_T^Q=\mathrm{max}_{\hat{X}}\mathcal{F}_T(\hat{X})$ with $\mathcal{F}_T(\hat{X})$ as in Eq.~(\ref{eq:fisher})~\footnote{Here, the notation is rather compact. More precisely, we mean the following: first, one needs to diagonalize the operator $\hat X$, whose eigenvectors are $\ket{x_s}$. Thus, the projection operators $\hat P_s = \ket{x_s}\bra{x_s}$ are defined. This is carried out for any observable $\hat X$. The Fisher information is then defined as a function of the set $\{\hat P_s\}$. Thus, the maximum over such a set for any $\hat X$ is sought. We note that one could also look for the optimal probe state, $\hat\rho_{\mathrm{P}}$, for a certain measurement, and therefore ask how the probe has to be prepared to attain the minimal uncertainty. Here, however, the state of the probe is determined by the interaction with the gas.}. 
It turns out that $\mathcal{F}_T^Q = \mathcal{F}_T(\hat{\Lambda}_T)$ with $\hat{\Lambda}_T$ being the symmetric logarithmic derivative operator (SLD)~\cite{Paris_IJQI09}
\begin{equation}
    \hat{\Lambda}_{T}\propto\cos(\varphi)\hat{\sigma}_\parallel+\sin(\varphi)\hat{\sigma}_\perp,\quad\tan(\varphi)=\frac{|\nu|(1-|\nu|)^2\partial_T\phi}{\partial_T|\nu|}.
    \label{eq:SLD}
\end{equation}
Here, $\hat{\sigma}_\parallel=\cos(\phi)\hat{\sigma}_x+\sin(\phi)\hat{\sigma}_y$, $\hat{\sigma}_\perp=\cos(\phi)\hat{\sigma}_y-\sin(\phi)\hat{\sigma}_x$ and $\nu=|\nu|e^{i\phi}$. In other words, the Fisher information retained in the projective measure of $\hat{\Lambda}_T$ on a two-level system is higher or equal than in the measurement of any other observable, and it equals the quantum Fisher information. It is for this reason that the protocol outlined below aims at measuring the observable represented by the symmetric logarithmic derivative and inferring the temperature and its uncertainty from it.\\ 
Before explaining the protocol, let us briefly discuss the estimation of the relative error on the measure of $T$. The quantum signal-to-noise ratio (QSNR), $\mathcal{Q}$, is defined via the quantum Fisher information by $\mathcal{Q}^2=T^2\mathcal{F}_T^Q$ and satisfies the following inequality
\begin{equation}
    \frac{\Delta T}{T}\geq \frac{1}{\mathcal{Q}\sqrt{N}},
    \label{eq:SNR_QSNR}
\end{equation}
which gives a lower limit of the attainable temperature uncertainty.\\
In the case of a two-level system, $\mathcal{F}_T^Q$ can be expressed in terms of the decoherence function $\nu(t)$~\cite{Nori_PRA13}. In polar coordinates, it reads
\begin{equation}
    \mathcal{F}_T^\mathcal{Q}=\frac{1}{1-|\nu|^2}\bigg(\frac{\partial|\nu|}{\partial T}\bigg)^2+|\nu|^2\bigg(\frac{\partial\phi}{\partial T}\bigg)^2=\mathcal{F}_T^\parallel+\mathcal{F}_T^\perp.
    \label{eq:QFI}
\end{equation}
Here, $\mathcal{F}_T^\parallel$ and $\mathcal{F}_T^\perp$ denote the contributions parallel and perpendicular to the Bloch vector, namely they correspond to the measurement of $\hat{\sigma}_\parallel$ and $\hat{\sigma}_{\perp}$, respectively.
More precisely, it means that if we would perform a projective measurement of the SLD in the eigenbasis of either $\hat{\sigma}_\parallel$ or $\hat{\sigma}_\perp$ we would obtain a Fisher information via the respective conditional probabilities $p(\Lambda_s^{\parallel, \perp}\vert T)$ of measuring the outcome $\Lambda_s^{\parallel, \perp}$ in one of the two eigenbasis given by either $\mathcal{F}_T^\parallel$ or $\mathcal{F}_T^\perp$. We note, however, that measuring the probe in those two eigenbasis implies that the gas temperature has to be known a priori, since the eigenvectors of $\hat{\sigma}_\parallel$ and $\hat{\sigma}_\perp$ depend on both angles $\phi$ and $\varphi$. In other words, one would need to set the measurement apparatus upon the gas temperature itself, which is our unknown. We shall come back to this point in the following paragraph.
\\[1ex]
\textit{Protocol. $-$} Let us describe briefly the main steps of the protocol sketched in Fig.~\ref{fig:protocol} in order to determine the expectation value of the SLD-operator. The atomic probe is initially prepared in the state $\ket{0}$. Therefore, the initial total density matrix is defined as: $\hat{\rho}=\ket{0}\bra{0}\otimes\hat{\rho}_\mathrm{B}(T)$. After a $\pi/2$-pulse, which takes the impurity-probe to the state $\ket{+}=(\ket{0}+\ket{1})/\sqrt{2}$, the system evolves for a time $t$ according to Eq.~(\ref{eq:blochvec}) with decoherence function given by Eq.~(\ref{eq:simplified_nu}). Then, a second $\pi/2$-pulse is performed with a certain phase $\theta$ and the energy is projectively measured. This kind of measurement yields an expectation value of the energy proportional to $\cos(\theta)\langle\hat{\sigma}_x\rangle+\sin(\theta)\langle\hat{\sigma}_y\rangle$. Accordingly to Eq.~\eqref{eq:SLD} as well as the definitions of $\hat{\sigma}_{\parallel}$ and $\hat{\sigma}_{\perp}$, one can determine $\langle \hat{\Lambda}_T \rangle$ by choosing $\theta=\phi+\varphi$. We repeat this procedure until the desired precision is reached. The number $N$ of independent repetitions realizes a measurement of the gas temperature with an error bounded from below by the right-hand-side of Eq.~\eqref{eq:SNR_QSNR}. In appendix~\ref{sec:Calculations} we provide some details on the aforementioned sequence of pulses. \\
A few comments are in order now. First, measuring the energy is an experimentally more feasible task than performing a measurement in the eigenbasis of $\hat{\sigma}_{\parallel}$ and $\hat{\sigma}_{\perp}$, which in addition rely on the unknown $T$. Second, the procedure outlined above is referred to estimation from the first moment in parameter estimation theory. Indeed, we estimate the gas temperature by measuring the expectation value of the SLD, i.e., the energy. Of course, the value of the phase $\theta$ depends on the gas temperature, but one would perform several repetitions of the procedure by varying $\theta$ in the Ramsey interferometer sequence. By assessing the Fisher information~(\ref{eq:fisher}) for various $\theta$, its maximum is reached for the actual gas temperature, i.e., our best estimation of $T$. Third, the corresponding uncertainty is given by $\Delta T = \Delta E (\partial \langle E \rangle / \partial T)^{-1}$ with $\Delta E^2$ being the variance of the energy. It can be shown that $\Delta T = 1/\sqrt{\mathcal{F}_T}$~\cite{refregier2012noise}. The standard deviation on the mean of the estimation is then given by $\Delta T/\sqrt{N}$, that is, the right-hand-side of Eq.~(\ref{eq:SNR_QSNR}). A crucial element of this discussion, however, is the determination of the conditional probability $p(\lambda_s\vert T)$ with $\lambda_s$ being the eigenvalues of $ \hat{\Lambda}_{T}$ and $s = \pm$. We shall come back to this point in Sec.~\ref{sec:Discussion}. \\
Finally, let us discuss the preparation of the initial product state $\hat{\rho}=\ket{0}\bra{0}\otimes\hat{\rho}_\mathrm{B}(T)$ that it is an assumption of the interferometric protocol, which is not so easibly realizeable as for a neutral impurity by tuning the impurity-bath interaction to zero, as we discussed at the beginning of Sec.~\ref{sec:System}. 
Towards this aim, we propose three solutions, the third one of which is otulined in more detail in the next paragraph. In the first solution we assume that the impurity is initially neutral and prepared in the internal state $\ket{0}$, whose interaction with the bath is tuned to zero by means of Feshbach resonances like in the original proposal of Ref.~\cite{Goold_PRL20}. Then, with a two-photon process as the one utilized in Ref.~\cite{pfau2018}, the ion is created and instantaneously (with respect to the system dynamics) its internal state is brought in the equatorial plane of the Bloch sphere and interactions with the gas take place upon the ion internal state. An alternative second solution relies on the utilization of Rydberg dressing of the atomic bath as suggested in Ref.~\cite{Negretti_PRA16}. More precisely, the ion is located at $\mu$m distances from the bath in order to ensure no interaction between it and the bath. Thus, the atoms in the bath are slightly coupled to Rydberg state in order to enhance the spatial range of the atom-ion interaction via the increase of the atomic polarizability of the bath. At this stage the equal superposition state of the ion internal states can be prepared. This solution, however, loses the in-situ character of the protocol, since the ion is positioned at a certain distance from the gas. 

\begin{figure}
    \centering
    \includegraphics[width=.4\textwidth]{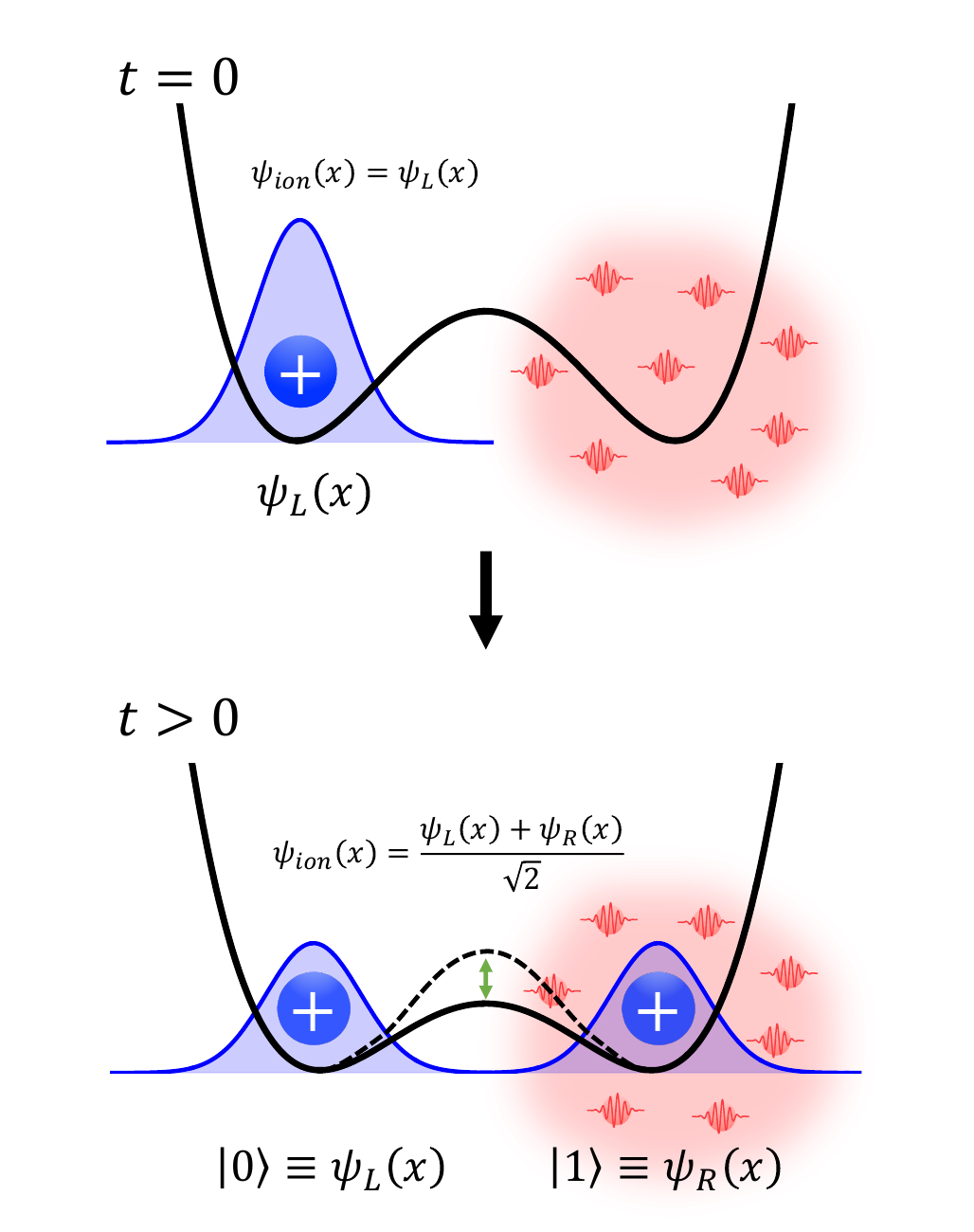}
    \caption{(Color online) Setup for a single internal state. Initially the ion is prepared in one of the two internal states, but the two-level system of the interferometric protocol is now represented by the ground states $\psi_L(x)$ and $\psi_R(x)$ of a double well potential. At $t=0$ the ion is prepared in the left well such that it does not interact in the bath. At $t>0$ the barrier is lowered such that tunnelling takes place and the ion is prepared in an equal superposition state of the left and right states. In this way, the ion's wavefunction component in the $\psi_R(x)$ state interacts with the bath and probing of its temperature can take place.}
    \label{fig:double_well}
\end{figure}

\textit{Setup for a single internal state. $-$}
We propose another solution for preparing the initial state $\hat{\rho}=\ket{0}\bra{0}\otimes\hat{\rho}_\mathrm{B}(T)$ that is particularly suitable if we aim at using only one ion internal state interacting with the gas. This solution mimics the case of the neutral impurity, where the interaction with the internal state $\ket{0}$ and the fermions is tuned to zero. Indeed, as we shall see in Sec.~\ref{sec:Results}, the use of a single internal state enables to attain an even higher sensitivity of the thermometer.\\
Specifically, the previously discussed interferometric protocol cannot be employed exactly in the same way, since the two-level system is reduced to a single internal state. To overcome this, we suggest to map the two-level system into the setup displayed in Fig.~\ref{fig:double_well}, where the states $\ket{0},\,\ket{1}$ are replaced by the left and right wavefunctions $\psi_L(x)$ and $\psi_R(x)$ of the ion in a double-well potential. Only the right well is immersed in the gas, letting the ion in state $\psi_L(x)$ to evolve freely without interaction with the gas. When the barrier is lowered, due to the tunnelling effect, the ion occupies the right well and at a precise time its state is described by an equal superposition of $\psi_L(x)$ and $\psi_R(x)$, thus mimicking a $\pi/2$-pulse. Immediately after that time, the barrier is raised again in order to suppress tunnelling and let the ion to probe the Fermi gas. When the QSNR maximum is attained, as we shall discuss in Sec.~\ref{sec:Results}, a second $\pi/2$-pulse is applied, namely the barrier is lowered again to allow tunnelling and thus to move the ion back to the left well. In such a way the previous interferometric protocol can be still applied, albeit with different type of measurements. \\
Let us remark that Ramsey interferometry with motional states of an atom in an optical lattice has been experimentally realized~\cite{hu2018} (also with the ground and first excited state of a quasi-1D condensate~\cite{vanFrank2014}). 
Additionally, the dynamics of Coulomb crystals in a double well potential has been theoretically investigated~\cite{Retzker2008,Klumpp2017}, while state-dependent optical potentials for trapped ions have been recently demonstrated in the laboratory~\cite{Weckesser2021}. The latter paves the way to engineer state-dependent potentials such as those created by microwave fields in atom chips~\cite{Treutlein2006,QIP:Boehi09}. These studies together with the possibility to employ optimal control methods~\cite{mueller2021decade} for steering the impurity dynamics appropriately corroborate the feasibility of the suggested scheme.

\textit{Decoherence and precision. $-$} To broadly understand how the thermometric precision can be affected by the parameters, we summarize some aspects of the decoherence dynamics. 
At short times, the decay of $|\nu|$ is due to collective excitations of the Fermi sea and the decoherence dynamics is essentially unaffected by the temperature. At large times, the dynamics is governed by low-energy excitations with temperature-dependent distribution and $|\nu|$ decays exponentially with rates depending on $T$~\cite{Schmidt_2018}. It is  intuitively clear that the probing time should be large enough in order to probe the gas temperature. According to Eq.~\eqref{eq:QFI}, however, high precision is achieved for large values of the derivative of $|\nu|$ with respect to $T$. Nonetheless, being the decay exponential at large times, it implies that long probing times decrease the quantum signal-to-noise ratio. For the same reason, a choice of the parameters that enhances the decoherence in the temperature-independent regime would result in less precise measurements. On this regard, we shall see in Sec.~\ref{sec:Results_0BS} how the effective interaction~\eqref{eq:convolution} affects the dependence on $k_\mathrm{F}a$ of the maximum of $\mathcal{Q}$ over time.

\section{Results}
\label{sec:Results}

To begin with, let us first provide a few pieces of information on the results we are going to discuss that will be helpful for the subsequent analyses. Our investigations will mostly focus on the case for which the ion is prepared only in one of the two internal states, since we aim at demonstrating the impact of the long-ranged atom-ion potential on the thermometer sensitivity (in Sec.~\ref{sec:Thermometry} an alternative interferometric protocol is suggested). In order to apply the protocol discussed in Sec.~\ref{sec:Thermometry}, however, we shall provide our findings corresponding to the scenario for which the ionic impurity is prepared in a superposition of the two internal states as well, where both states interact with the gas. Hence, unless explicitly indicated, the interaction parameter $k_\mathrm{F}a$ refers to the case with the single interacting state $\ket{s}$ with $s = 0$ or 1 and the atom-ion scattering length is assumed to be $a\simeq -R^\star$~\footnote{The choice of a negative scattering length is motivated by recent experimental investigations~\cite{Gerritsma_NAT20,Schaetz_NAT21}.}. 
We shall explore different scenarios of the regularized atom-ion polarization potential (also shown in Fig.~\ref{fig:potentials}), for which the absence or presence of one or more bound states is assumed. This will enable us to investigate the impact of bound states on the dynamics of the probe and, most importantly, on the precision of the temperature estimation.
Let us remark that the choice of a negative scattering length motivates the scenario without bound states. In fact, a negative scattering length implies that the corresponding state is deeply bound and that it has a rather small size. Hence, if recombination timescales are long enough, it is unlike that such states are populated. We note that in a different context than ours, precisely in this regime of interactions and time scales phenomena such as attractive polarons have been predicted~\cite{Gregory2021}. We shall turn back to this matter in Sec.~\ref{sec:Discussion}.

As far as the interaction parameter $k_\mathrm{F}a$ is concerned, we note that, differently from the neutral impurity case where the $s$-wave scattering length is the only parameter characterizing the two-body interaction, for the long-ranged atom-ion polarization potential the reliance of $k_\mathrm{F}a$ on the scattering length is less trivial. Indeed, its typical length scale, $R^\star$, is comparable to the mean inter-particle distance and the system properties depend not only on $a$ and the effective range of the two-body potential, but also rely on the presence of the long-range tail of the interaction. Hence, while in the neutral case a specific choice of the product $k_\mathrm{F}a$ can be obtained for a fixed gas density by a single value of the scattering length, in the atom-ion scenario the same value of the scattering length can result in different two-body potentials leading to different system properties.\\ 
In Table~\ref{tab:values} we report the Fermi time and temperature for different atom-ion pairs for two values of the interaction parameter $k_\mathrm{F}=0.5/R^\star$ and $k_\mathrm{F}=1.5/R^\star$, from which we obtain the corresponding gas densities \footnote{As it can be seen, there is a rather large variation of the two time and energy scales for three exemplary atom-ion pairs. In particular, the Fermi temperature for the pair $^6$Li – $^{174}$Yb$^+$ is the largest, reflecting the fact that a lighter atom with a heavy ion are also easier to cool down to the $s$-wave collisional regime, while for the pair $^{40}$K – $^{174}$Yb$^+$ with the largest $R^\star$ (due to the larger atomic polarizability) it is more challenging to attain the quantum degeneracy. Furthermore, larger gas densities reduce the coherence time $\tau_\mathrm{F}$ which has the smallest value for lighter atoms, that is, for the atom-ion pair $^6$Li – $^{174}$Yb$^+$.}.\\
\begin{table*}
    \centering
    \setlength{\extrarowheight}{1ex}
    \setlength{\tabcolsep}{2ex}
    \begin{tabular}{rccc}
        \hline\hline
        \textbf{Atom – Ion} $\boldsymbol{(R^\star}\,(\mathrm{nm})\boldsymbol{)}$ & \textbf{Mean Density} $(\mathrm{cm}^{-3})$ & $\boldsymbol{\tau_\mathrm{F}}$ $(\mathrm{\mu s})$ & $\boldsymbol{T_\mathrm{F}}$ $(\mu \mathrm{K})$ \\[1ex]
        \hline\hline
        \multirow{2}{10em}{$^6$Li – $^{174}$Yb$^+$ ($69.77$)} & $6.2\times 10^{12}$ & $3.7$ & $2.07$ \\ & $1.7\times 10^{14}$ & $0.41$ & $18.64$\\
        \hline
        \multirow{2}{10em}{$^{40}$K – $^{174}$Yb$^+$ ($219.24$)} & $2.0\times 10^{11}$ & $240$ & $0.032$\\ & $5.4\times 10^{12}$ & $26$ & $0.29$ \\
        \hline
         \multirow{2}{10em}{$^{40}$K – $^{40}$Ca$^+$ ($171.92$)} & $4.15\times 10^{11}$ & $145$ & $0.053$ \\ & $1.1\times 10^{13}$ & $16$ & $0.47$\\
        \hline\hline
    \end{tabular}
    \caption{Most relevant physical quantities for some atom-ion species: $k_\mathrm{F}=0.5/R^\star$ (top row) and $k_\mathrm{F}=1.5/R^\star$ (bottom row).}
    \label{tab:values}
\end{table*}

\subsection{Static ion approximation}
\label{sec:Results_0BS}
\begin{figure*}
    \centering
    \includegraphics[width=.4\textwidth]{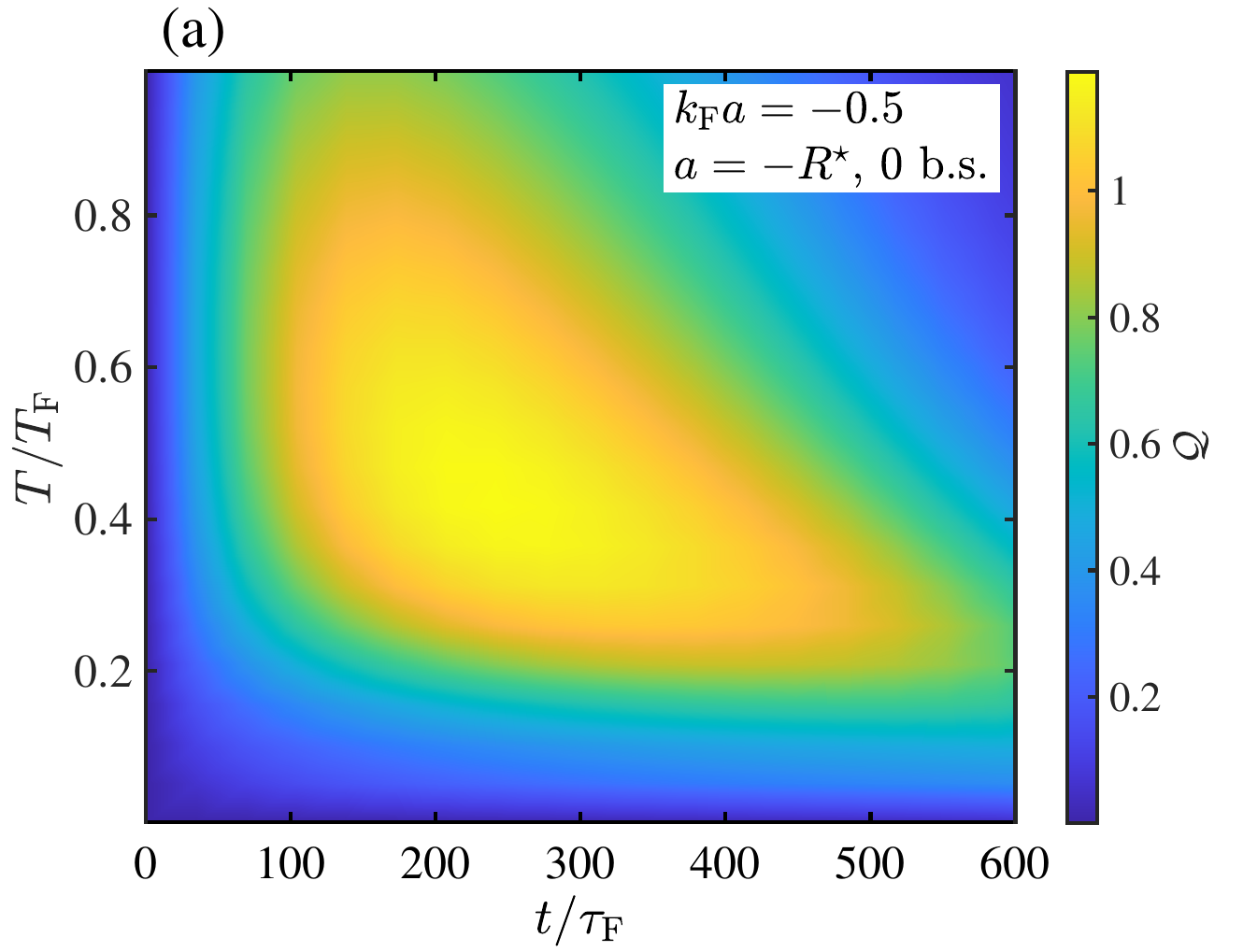}
    \includegraphics[width=.4\textwidth]{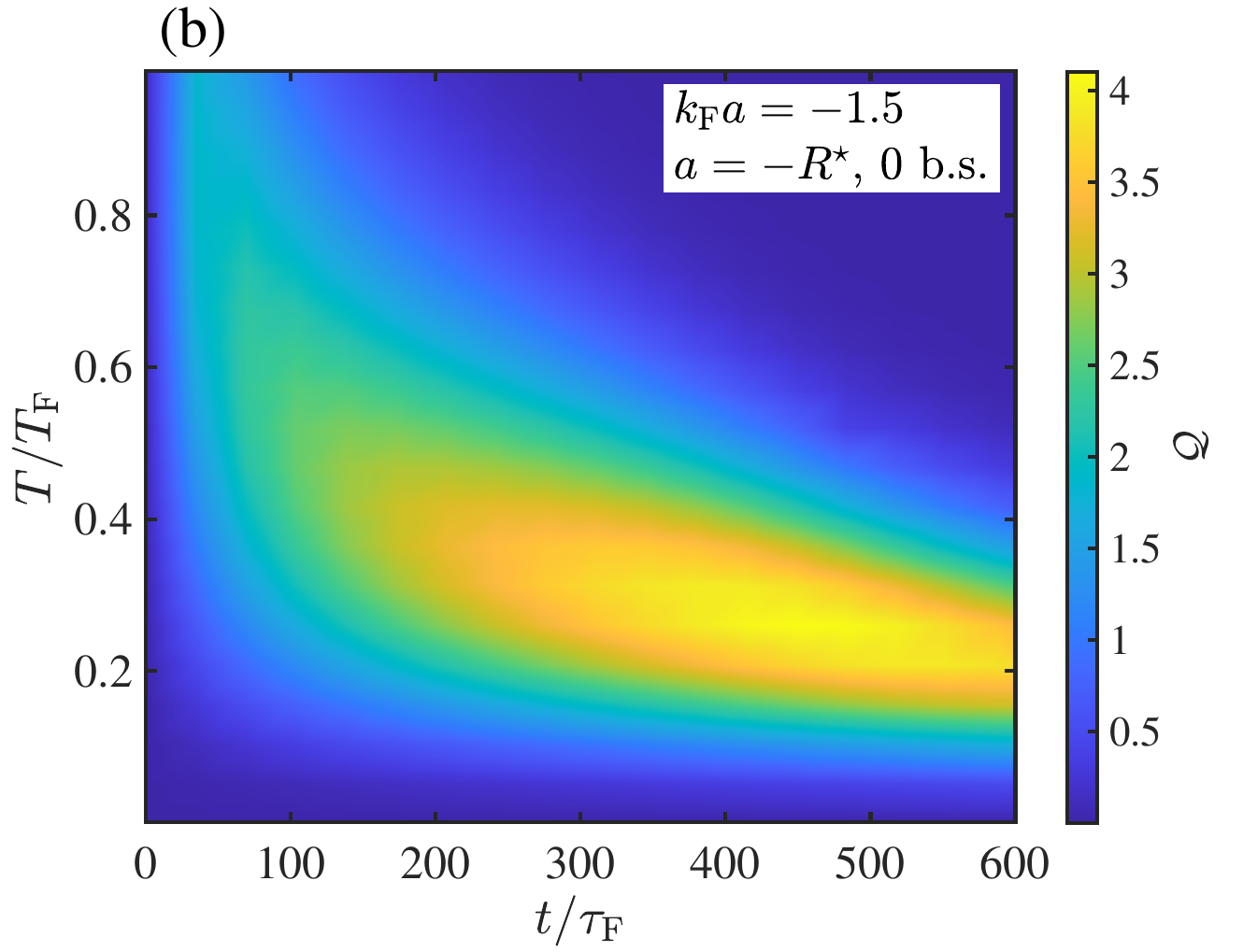}\\
    \includegraphics[width=.4\textwidth]{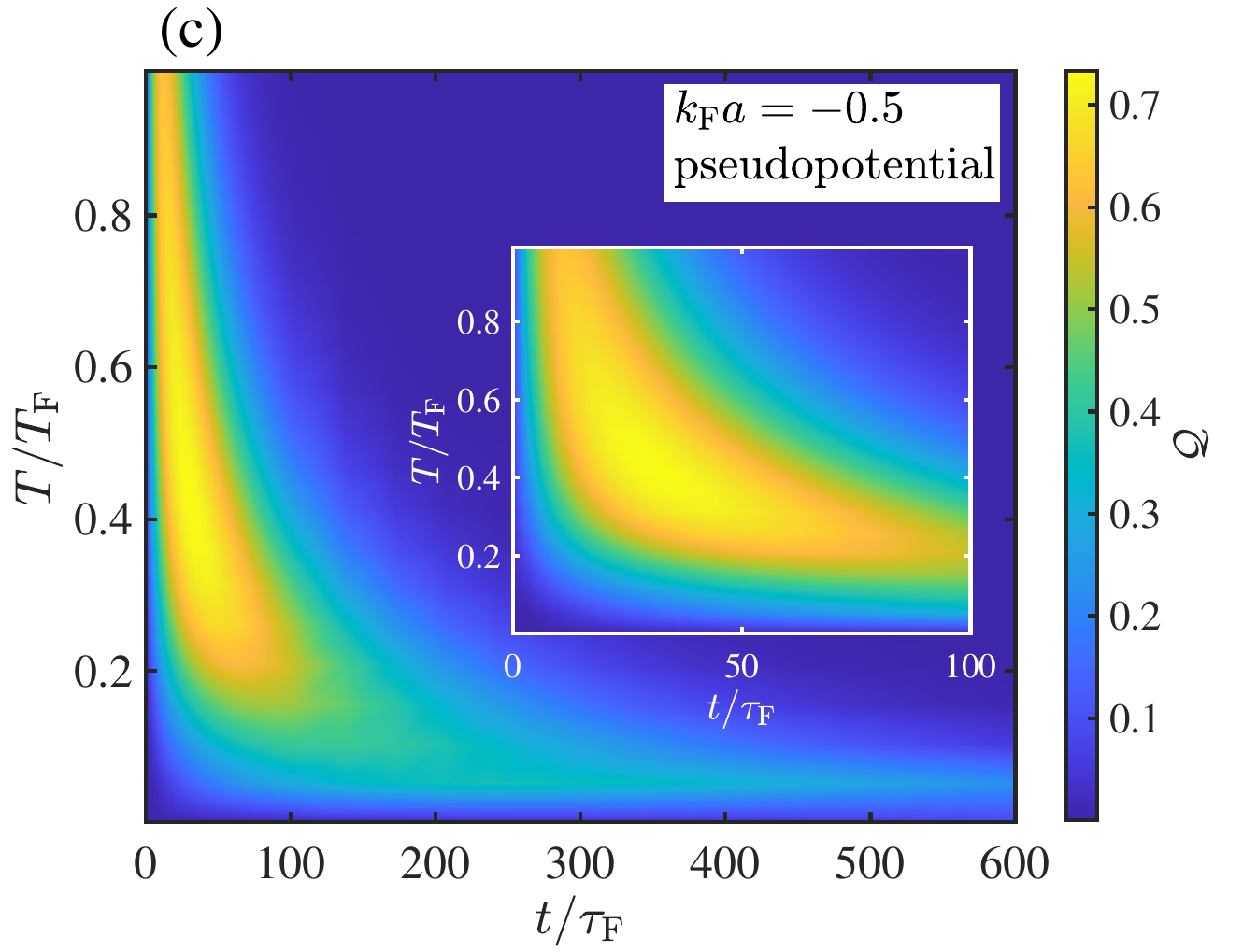}
    \includegraphics[width=.4\textwidth]{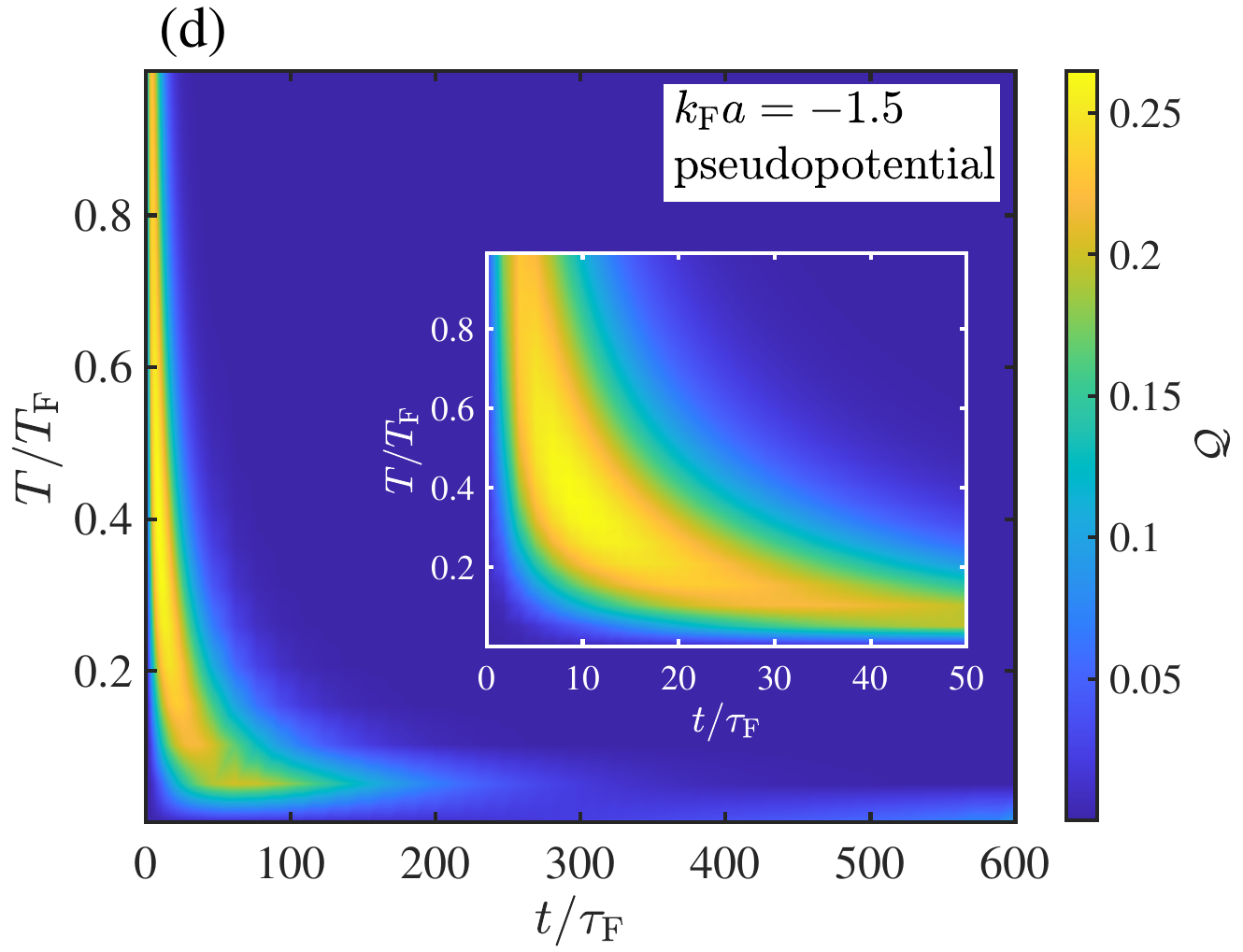}
    \caption{\label{fig:QSNR2D_NBS} Quantum signal-to-noise ratio for $k_\mathrm{F}a=-0.5$ and $k_\mathrm{F}a=-1.5$ in the static impurity approximation. Panels (a) and (b) for an ion with an atom-ion polarization potential without bound states, whereas panels (c) and (d) for a neutral impurity with impurity-bath interaction given by the pseudopotential.}
\end{figure*}
\begin{figure}
    \centering
    \includegraphics[width=.4\textwidth]{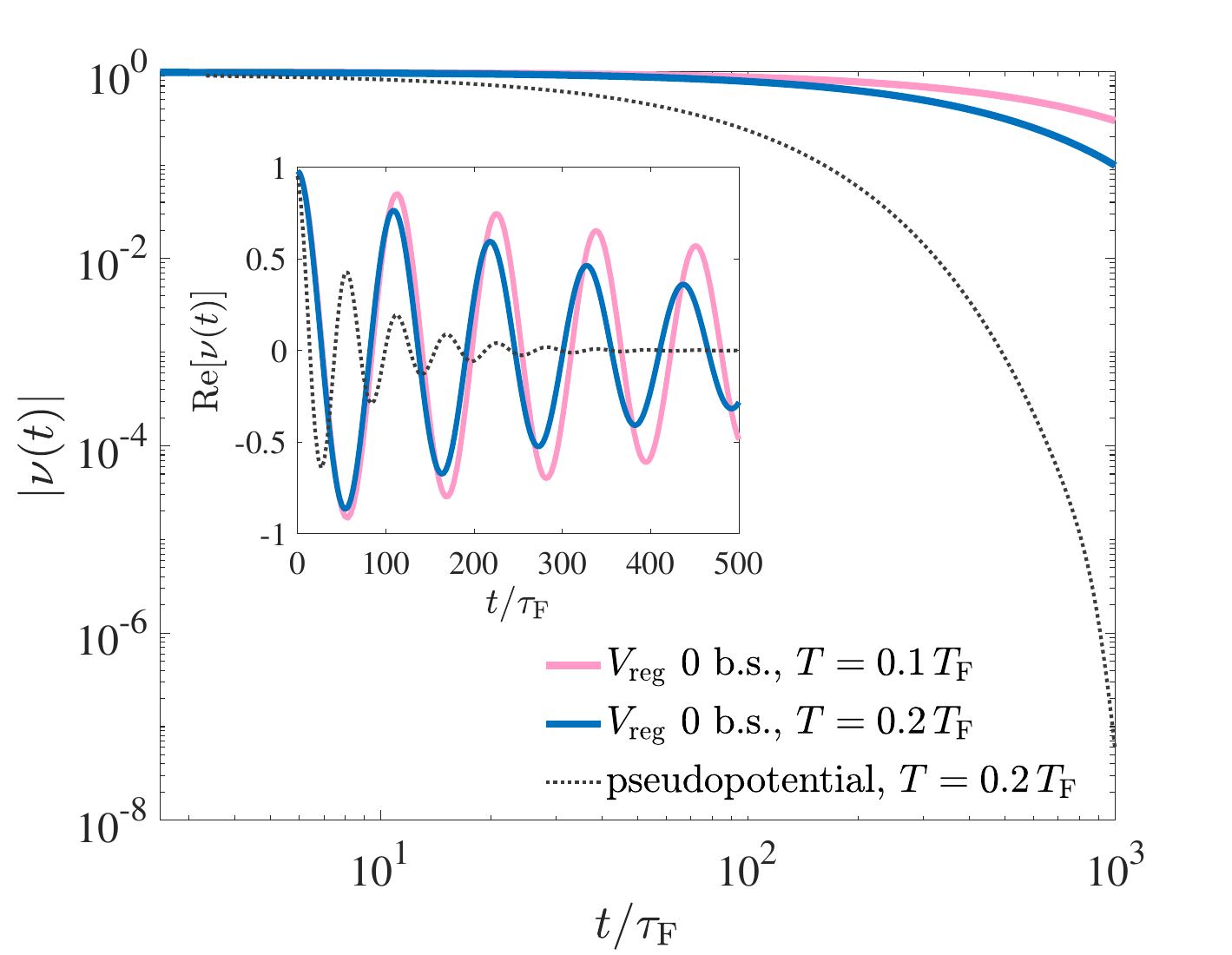}
    \caption{(Color online) Decoherence function for $k_\mathrm{F}a=-0.5$ for a static impurity. Solid line for the atom-ion polarization potential without bound states and two different temperatures (see legend), while the dotted line for the pseudopotential. In the main panel the time dependence of $\vert \nu(t) \vert$ is shown, while in the inset its real part is displayed.}
    \label{fig:decoherence}
\end{figure}
\begin{figure}
    \centering
    \includegraphics[width=.48\textwidth]{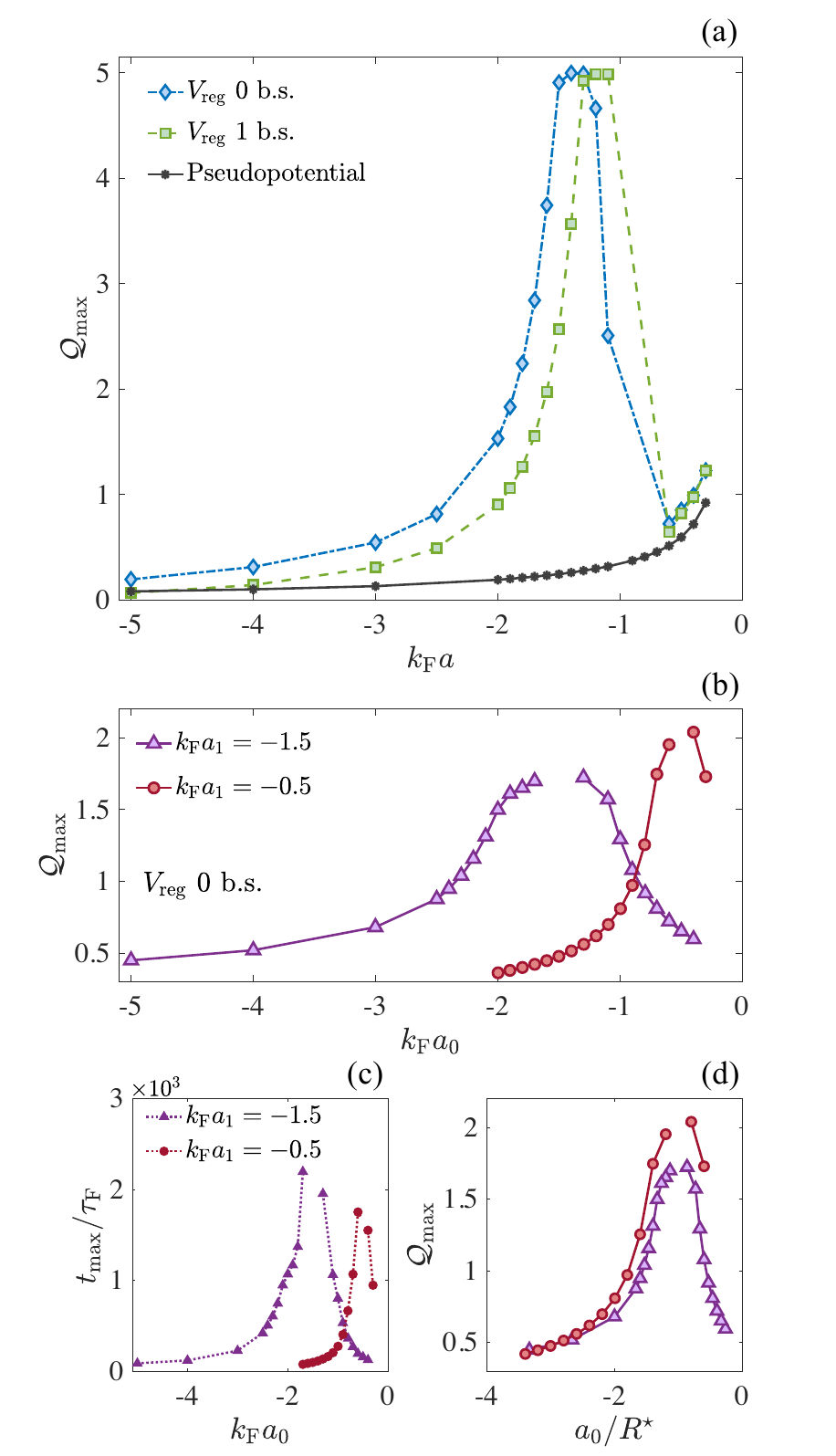}
    \caption{(Color online)
    Panel (a): $\mathcal{Q}_\mathrm{max}\equiv\mathcal{Q}(t_\mathrm{max})$ for different values of $k_\mathrm{F}a$ at $T=0.2\,T_\mathrm{F}$ and for an internal ion state only (e.g., the state $\ket{1}$). Panel (b): $\mathcal{Q}_\mathrm{max}$ for different values of $k_\mathrm{F}a_0$ (i.e., state $\ket{0}$) at $T=0.2\,T_\mathrm{F}$ and fixed $k_\mathrm{F}a_1=-1.5\,R^\star$ (purple triangles) and $k_\mathrm{F}a_1=-0.5\,R^\star$ (red circles), i.e., state $\ket{1}$. Panel (c): values of $t_\mathrm{max}$ corresponding to panel (b). Panel (d): same data of panel (b) plotted against $a_0$.  All the lines are merely a guide to the eye. Their discontinuity in panels (b), (c) and (d) indicates that the thermometer does not work when $k_\mathrm{F}a_0=k_\mathrm{F}a_1$ (see text).}
    \label{fig:Qmax}
\end{figure}
We shall first consider the case with zero bound states (dot-dashed line in Fig.~\ref{fig:potentials}). This will allow us for a direct comparison with the pseudopotential for neutral atoms and thus highlight the effect of the long-range character of the polarization potential more clearly. For the same reason, we shall first consider the case for which the ion probe is localized at $r=0$ with $|\chi(\mathbf{r})|^2=\delta(\mathbf{r})$. Note that this is how the static ion approximation has to be intended and it is achieved either by confining the ion in a very tight trap or by considering the mass of the ion to be much larger than the mass of the atoms in the gas, i.e., infinite mass limit. To compare directly with the neutral impurity case, our analysis begins with the case where only one of the two internal states interacts with the bath. The parameters for this regularized potential are $b\simeq 0.0023\,R^\star$ and $c\simeq 0.4878\,R^\star$ that correspond to $a\simeq-R^\star$~\footnote{In presence of two-body bound states the parameters of the regularized atom-ion polarization potential assume the following values: $b\simeq 0.07018\,R^\star$ and $c\simeq 0.1455\,R^\star$ for one supported bound state, whereas $b\simeq 0.09\,R^\star$ and $c\simeq 0.0718\,R^\star$ for two bound states.}.
We finally remark that the QSNR is calculated via the relation $\mathcal{Q}^2=T^2\mathcal{F}^\mathcal{Q}_T$, where the quantum Fischer information is obtained from Eq.~\eqref{eq:QFI}.
\\[1ex]
\textit{Temperature and time dependence of QSNR. $-$} We consider $k_\mathrm{F}a=-0.5$ and $k_\mathrm{F}a=-1.5$ in the case with one internal state and we start by analyzing the temperature and time dependence of the quantum signal-to-noise ratio. These two choices result in values of the mean density and Fermi time of typical quantum gas experiments (see also table~\ref{tab:values}). As we can observe in Fig.~\ref{fig:QSNR2D_NBS}, the temperature dependence is only slightly affected by the two different types of potentials (atom-ion polatization interaction vs. pseudopotential) for both of the two values of the interaction parameter. Indeed, the yellow (i.e., bright) region corresponding to the maximum of $\mathcal{Q}$ appears roughly in the same range of the Fermi temperature ratio [$\sim 0.4$, see panels (a) and (b) vs. panels (c) and (d)]. The time dependence of $\mathcal{Q}$ for the two $k_\mathrm{F}a$, however, is significantly modified: the region of highest precision, i.e. higher  $\mathcal{Q}$, it is extended to longer probing times for a charged impurity [panels (a) and (b)], while it occurs at earlier times and for a smaller time window for the case of a neutral impurity [panels (c) and (d)]. This behavior suggests that $V_\mathrm{reg}$ suppresses the decay of $|\nu|$, as it also clearly showed in Fig.~\ref{fig:decoherence}. Moreover, the shift of the yellow region to longer times also suggests that the potential takes more time to bring the system in the regime where the dynamics is governed by particle-like excitation, which are more sensitive to the gas temperature. 
Concerning that point, Fig.~\ref{fig:decoherence} shows clearly that the long time behaviour of the decoherence function is more affected by the gas temperature as well as that the larger the temperature, the stronger the effect on the decoherence function. As a consequence, it results in a better temperature probe at long times.\\
Finally, Fig.~\ref{fig:QSNR2D_NBS} shows that the use of an ion probe is more favorable in some circumstances. Indeed, in the case of low densities [panels (a) and (c)], i.e. small $\vert k_\mathrm{F}\vert$, a limited gain in the thermometer sensitivity can be obtained, but for longer interrogation times. On the other hand, at large values of $\vert k_\mathrm{F}a\vert$ [panels (b) and (d)], that is, high gas densities, i.e. large $\vert k_\mathrm{F}\vert$ for the polarization potential, the use of an ion instead of a neutral impurity particle significantly enhances $\mathcal{Q}$ (up to 5 times more at the maximum), and therefore the temperature sensor accuracy, in a large part of the considered time range. Hence, such findings demonstrate that using ionic impurities for sensing the temperature of a Fermi gas result in a superior thermometer performance.\\[1ex]
\textit{Interaction dependence of the maximal QSNR at fixed T. $-$} At fixed $T=0.2\,T_\mathrm{F}$, Fig.~\ref{fig:Qmax}(a) shows that the monotonous behavior of $\mathcal{Q}_\mathrm{max}=\mathrm{max}_t\{\mathcal{Q}(t)\}$ in the case of the neutral impurity (grey asterisks) is not reproduced with the ion, where instead a peak appears (blue diamonds). In order to understand the onset of the latter, one can observe that the two potentials give a similar dependence on $k_\mathrm{F}a$ in the extremal regions of the considered range of Fig.~\ref{fig:Qmax} (i.e., small and large magnitudes of $\vert k_\mathrm{F}a \vert$). 
Indeed, the dependence on $k_\mathrm{F}a$ can be mapped to a dependence on the mean inter-particle distance of the bath particles $\Bar{d}=(\bar{n})^{-1/3}\propto1/|k_\mathrm{F}a|$~\footnote{Here, the overbar indicates that the length has been rescaled with respect to $a$.}. When $\bar{d}$ is sufficiently large, i.e. $\bar{d}\gtrsim8\,R^\star$ corresponding to  $k_\mathrm{F}a\gtrsim-0.5$ and small densities, the two potentials give similar results. In this regime, the rate of collisions affecting the ionic probe is small and the decoherence function $\nu(t)$ decays slowly compared to $\tau_\mathrm{F}$ (see also Fig.~\ref{fig:decoherence}). This allows the ion to probe the bath for a longer time, thus resulting in increasing $\mathcal{Q}_\mathrm{max}$ values for decreasing values of $|k_\mathrm{F}a|$. On the contrary at large densities, i.e. $\bar{d}\lesssim R^\star$ and $k_\mathrm{F}a\lesssim-4$, the rate of collisions between the particles of the bath and the strongly repulsive core of the potential is higher. Both with the regularized pseudopotential and $V_\mathrm{reg}$, when $|k_\mathrm{F}a|$ increases, the decoherence function $\nu(t)$ decays more rapidly (not shown) and the values reached by $\mathcal{Q}_\mathrm{max}$ are lower
\footnote{Let us underscore that the interaction parameter $|k_\mathrm{F}a|$ has to be interpreted differently for the neutral and charged impurity. Indeed, for the ionic impurity we keep fixed the scattering length, while we vary the wave-vector $k_\mathrm{F}$, i.e., the gas density. Conversely, when the impurity is neutral, although the value of $|k_\mathrm{F}a|$ can in principle correspond to any of the possible combinations of the two terms, one has to consider a fixed gas density, and hence $k_\mathrm{F}$, and a varying scattering length. Otherwise, at large magnitudes of the interaction parameter, the applicability of the pseudopotential would be invalided, as $\bar{d}$ would be comparable or even smaller than the effective range of the van der Waals interaction.}.
When the probe is an ion, an intermediate regime can be identified where the balance between interactions with the repulsive core and the attractive well of $V_\mathrm{reg}$ strongly suppresses the decay of $\nu(t)$, and therefore allowing to attain much higher values of the quantum signal-to-noise ratio. 
Besides this, we attribute the enhanced precision to the accumulated phase given by the long-range polarization potential, i.e., the relative phase proportional to the integral of the atom-ion potential over the density perturbations of the bath because of the presence of the impurity. Since the atom-ion potential is long-range, it will collect more of those perturbations. However, as in the case of low densities, a large $\mathcal{Q}$ requires a long probing time which could affect the effectiveness of the protocol. 
Note that the attractive region of the potential with no bound states is not visible in Fig.~\ref{fig:potentials} (blue dot-dashed line), because it is considerably more shallow than the other two cases. Similar results are obtained when the atom-ion polarization potential supports one bound state [green squares in Fig.~\ref{fig:Qmax}(a)] or two bound states (not shown) with the only difference that a slight shift of the maximum of $\mathcal{Q}_\mathrm{max}$ occurs. This indicates that the phenomenon does not rely on the number of bound states, but it is rather a peculiarity of the long-range character of the two-body interaction.
As already mentioned, the most relevant case for the application of the protocol explained in Sec.~\ref{sec:Thermometry} is the one where the two internal states interact with the gas, whose result is displayed in Fig.~\ref{fig:Qmax}(b), (c) and (d). To understand the plots, it is necessary to quickly explain the meaning of the interaction parameters $k_\mathrm{F}a_1$ and $k_\mathrm{F}a_0$. The choice of the value $k_\mathrm{F}a_1=-1.5\,R^\star$ or $k_\mathrm{F}a_1=-0.5\,R^\star$ (purple triangles and red circles respectively), with $a_1\simeq-R^\star$ the scattering length relative to the state $\ket{1}$, fixes the density of the bath. Consequently, the variation of $k_\mathrm{F}a_0$ corresponds to the tuning of the atom-ion scattering length $a_0$ of the state $\ket{0}$. In this way, different values of the interaction parameter $k_\mathrm{F}a_0$ are obtained without changing the density of the bath. Note that the thermometer does not work when $k_\mathrm{F}a_0=k_\mathrm{F}a_1$, since the two potentials are identical and therefore the induced dephasing dynamics is the same. This is can be also recongnized in Eq.~\eqref{eq:matrix_element} and it is indicated in the panels (b)-(d) of Fig.~\ref{fig:Qmax} by the discontinuity of the connecting line. The plot in Fig.~\ref{fig:Qmax}(b) shows that the non-monotonous behavior that we attribute to the long-range of the potential is preserved and that a proper choice of the parameters can lead to a higher sensitivity. Other than the enhancement of the sensitivity, a shift in the position of the peaks is observed, depending on the value of $k_\mathrm{F}a_1$. This is better understood by observing the QSNR as a function of the scattering length $a_0$ instead of the interaction parameter $k_\mathrm{F}a_0$. As shown in Fig.~\ref{fig:Qmax}(d), the dependence of the QSNR on  $a_0$ is similar for the two cases and the peak appears around $a_0=-R^\star$. 
Finally, we note that the time at which each $\mathcal{Q}_\mathrm{max}$ occurs [see Fig.~\ref{fig:Qmax}(c)] is slightly longer than the case of the single internal state. The latter is not shown, but it does not exceed $10^3\tau_\mathrm{F}$ for the values in Fig.~\ref{fig:Qmax}(a). This difference can be attributed to the cancellation between the accumulated phases in the elements of the matrix from which the decoherence function is calculated [see Eq.~\eqref{eq:decoherence} and Eq.~\eqref{eq:matrix_element}]. On the other hand, the peaks of $\mathcal{Q}_\mathrm{max}$ are smaller than the single internal state case, but still larger than the neutral impurity probe. We attribute this reduction of sensitivity to the previous argument of the accumulated phases around the density perturbations, that is, the two states compensate partially each other.\\

\subsection{Finite ion density distribution}
\label{sec:Results_LowT}
\begin{figure}
    \centering
    \includegraphics[width=.4\textwidth]{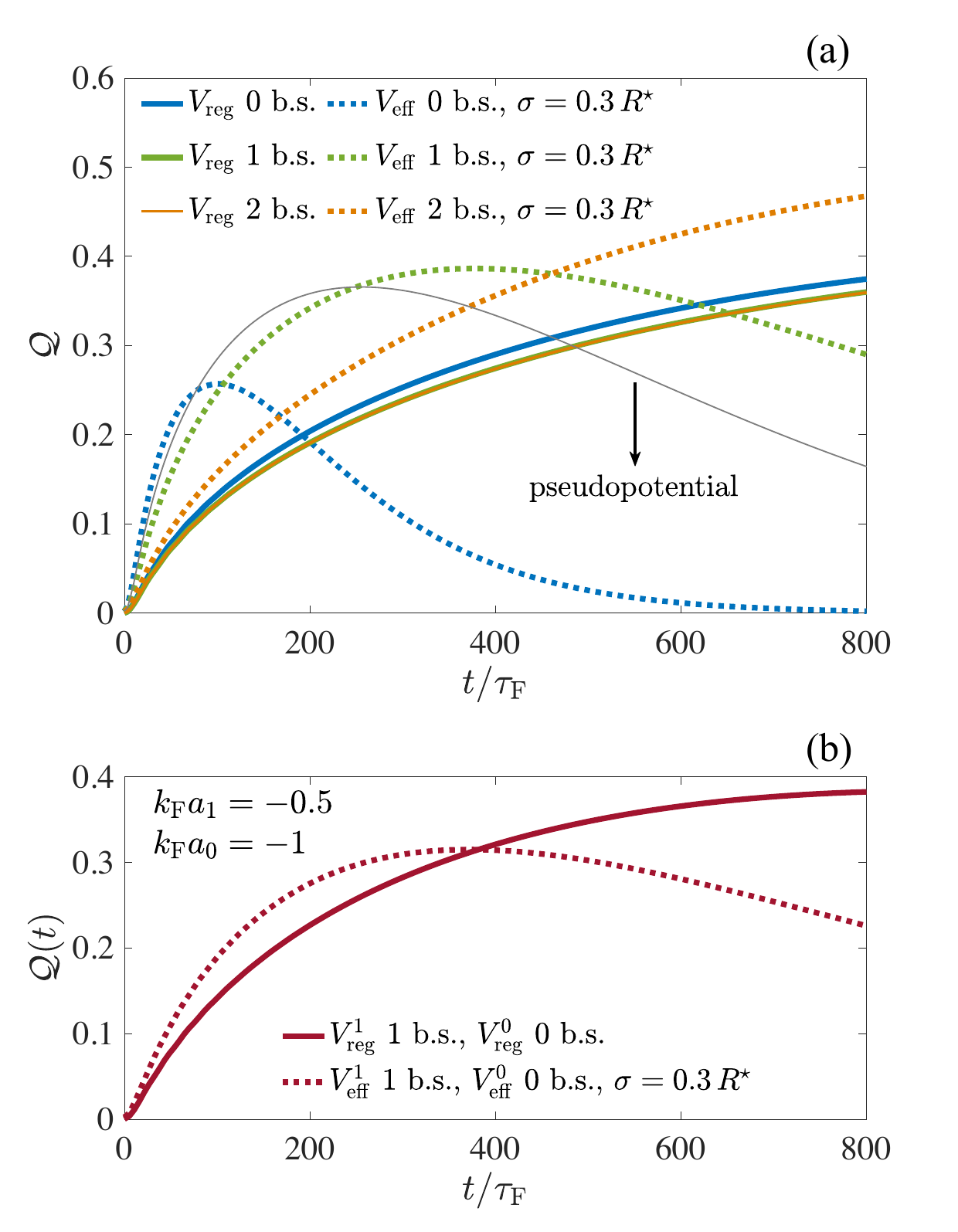}
    \caption{(Color online) Quantum signal-to-noise ratio as a function of the probing time for $T=0.05\,T_\mathrm{F}$. Panel (a): single state with $k_\mathrm{F}a=-0.5$. Panel (b): two states with $k_\mathrm{F}a_1=-0.5$ and $k_\mathrm{F}a_0=-1$. Solid lines represent the localized impurity case; dotted lines correspond to the impurity with Gaussian spatial density.}
    \label{fig:QSNRs}
\end{figure}
\begin{figure}
    \centering
    \includegraphics[width=.4\textwidth]{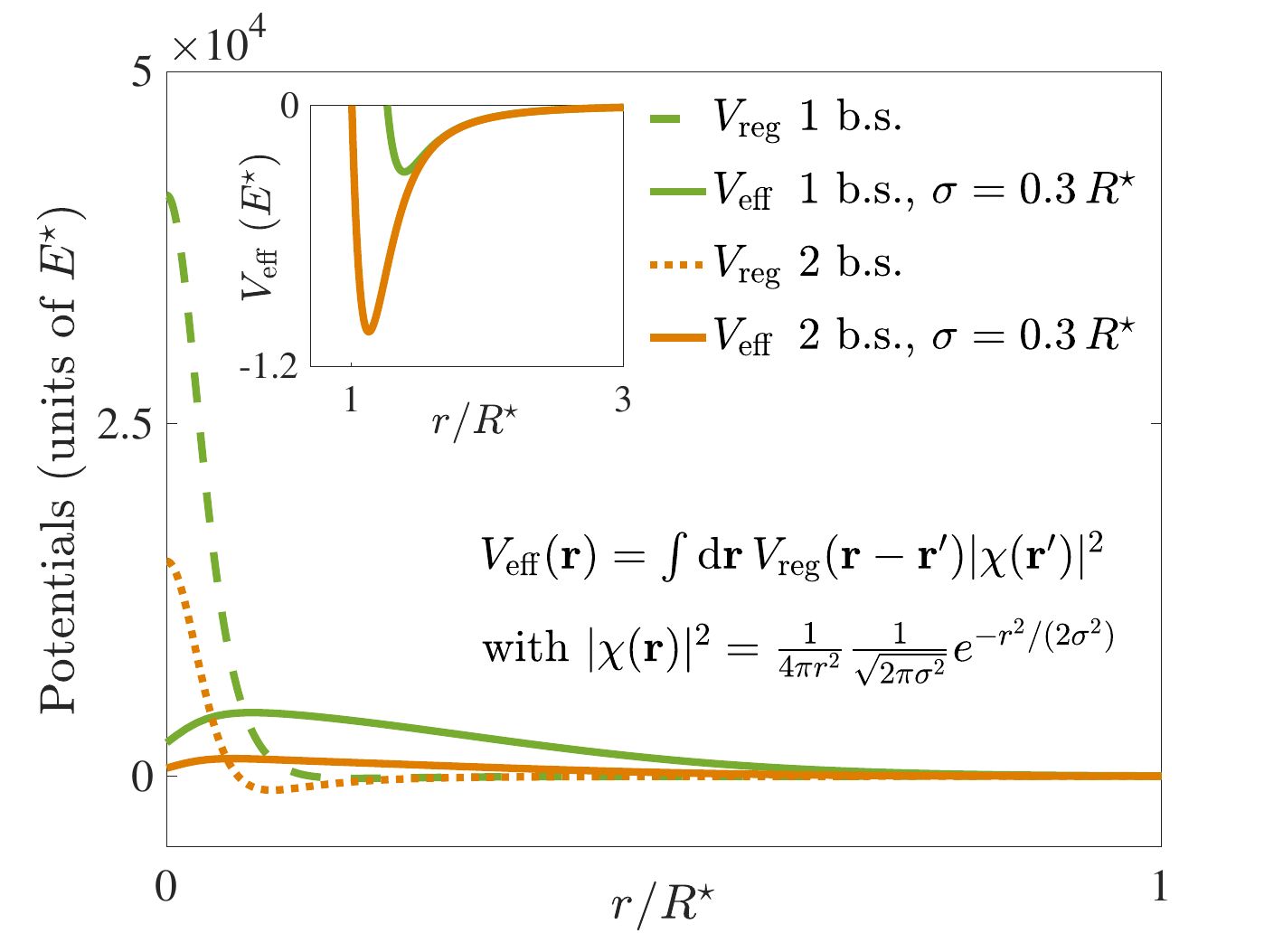}
    \caption{(Color online) Regularized atom-ion potentials with one (green line) and two (orange line) bound states together with their corresponding effective potentials (see also Sec.~\ref{sec:System}). The inset shows the attractive region of the effective potentials. Note that the case without bound states is not shown, because both the energy and the atom-ion separation are of a rather different scale compared to those shown in the picture. Specifically, one has to consider much larger separations and no additional novel feature would have been manifested.}
    \label{fig:effective_potentials}
\end{figure}
We now investigate the impact of a spatial density distribution of the ion as that obtained when the ion is confined in a trap of finite width. In particular, we choose a Gaussian distribution, as it approximates the time-averaged distribution of the ion subject to micromotion in the ground state of a Paul trap (secular approximation) or the ground state in a deep optical dipole trap. The width of the Gaussian distribution is defined as $\sigma = \sqrt{\hbar/M\omega}$ with $M$ the ion mass and $\omega$ the frequency of the trap. A trap width of $\sigma=0.3\,R^\star$ corresponds to a trap frequency $\omega/(2\pi) \simeq 133$ kHz for $^{174}$Yb$^+$ and $\omega/(2\pi) \simeq 95$ kHz for $^{40}$Ca$^+$ (see also table~\ref{tab:values}).
In Fig.~\ref{fig:QSNRs} we show exemplary the time evolution of the quantum signal-to-noise ratio for the case of 
$T=0.05\,T_\mathrm{F}$. The thick solid lines in Fig.~\ref{fig:QSNRs}(a) show the result for one single internal state with $k_\mathrm{F}a=-0.5$ in the static ion approximation with zero (blue line), one (green line) and two (orange line) bound states. The latter two are essentially superimposed to each other (almost indistinguishable in the plot). As it can be seen, apart from a slight difference between the case of zero bound states and those with a finite number of bound states, the QSNR does not exhibit any reliance on the number of bound states when the ionic probe's distribution is delta-shaped. Moreover, $\mathcal{Q}$ assumes large values only at long times, where the impact of detrimental effects such as three-body recombination or reduced trap lifetime, especially for optical-based trap technology, are more likely. On the other hand, when the Gaussian distribution of the probe is considered, the convolution between the latter and the regularized potential results in an effective potential $V_\mathrm{eff}$ [see also Eq.~(\ref{eq:convolution})]. Such an effective potential has different characteristics compared to $V_\mathrm{reg}$ and it substantially modifies the dynamics of $\mathcal{Q}$. Indeed, as it is also visible in Fig.~\ref{fig:effective_potentials}, the attractive region becomes more shallow and the repulsive core gets flattened and broadened. Interestingly, a reliance on the number of bound states is manifested (see dotted lines in Fig.~\ref{fig:QSNRs}), especially for zero and one bound states, where a maximum at short times is displayed, which enables to attain higher sensitivities of the ionic temperature sensor at  shorter times and thus to reduce the impact of undesired effects. On the other hand, the situation with two bound states resembles the case of the static ion limit, albeit attaining larger $\mathcal{Q}$ values at longer times (see orange dotted line). This shows that the deeply bound states have a marginal impact on the sensor performance. A similar behavior is shown in Fig.~\ref{fig:QSNRs}(b) where two internal states are considered. In this case, the interaction parameters are fixed to $k_\mathrm{F}a_0=-1$ and $k_\mathrm{F}a_1=-0.5$, and the regularized potentials support zero and one bound states for $\ket{0}$ and $\ket{1}$, respectively. For times shorter than $t\simeq400\,\tau_\mathrm{F}$, a finite spatial density (dotted line) results in slightly higher values of the QSNR with respect to the case with the static ion (solid line). Although the difference is not remarkable, Fig.~\ref{fig:QSNRs}(b) confirms that, in general, the ion finite spatial density allows to improve the sensitivity at shorter times, even though the sensor performance is worse compared to the single internal state. Hence, our analysis indicates that the trap frequency $\omega$ can be used as a ``knob'' to enhance the signal-to-noise ratio at short times, which is particularly relevant in view of spin relaxation~\cite{ratschbacher2013}.
\\ 
We finally remark that other strategies to determine the temperature of an ultracold Fermi gas were proposed. In particular, Lous \textit{et al.} showed in Ref.~\cite{Grimm_PRA17} that a temperature $T\simeq 0.06\,T_\mathrm{F}$ of a $^6$Li gas can be estimated with a $10\%$ error by using a Bose-Einstein condensate of $^{41}$K atoms as a probe. The strategy is based on the thermalization between the two atomic species and on the estimation of the condensate fraction. Their aim, however, was to propose an experimental technique to determine the lowest possible temperature of the fermionic gas specifically for their setup. Here, on the other hand, we provide a theoretical estimation of the interrogation time, i.e., when the maximum achievable QSNR is attained, and the number of measurements that are needed to obtain a comparable error. For instance, a precision $\Delta T/T=0.1$ of a measured temperature $T=0.05\,T_\mathrm{F}$ can be obtained with $N\approx625$ repetitions on a time of the order of a few milliseconds ($\sim 1000\tau_{\mathrm{F}}$) for an atom-ion pair $^6$Li – $^{174}$Yb$^+$, where the number of repetitions $N$ can be reduced by employing more ions as sensors at the same time. We underscore again, however, that the experimental achievement of such a measurement relies on the capability of performing the protocol outlined in Sec.~\ref{sec:Thermometry}. 

\section{Discussion}
\label{sec:Discussion}
The results we exposed in the previous section show that an impurity ion, whose interaction with the bath is described by the long-range atom-ion polarization potential, yields a higher quantum signal-to-noise ratio than a neutral impurity in an experimentally accessible parameter regime. Nonetheless, a few remarks on the interferometric protocol as well as on inter-particle collisions are in order. \\[1ex]
\textit{Interferometric protocol. $-$} The protocol outlined in Sec.~\ref{sec:Thermometry} aims at determining the symmetric logarithmic derivative via the projective measurement of the energy. To this end, one performs a series of independent experimental runs by collecting data as $\{\xi_k\}_{k=1}^N$ with $\xi_k$ being the outcome of the measurement of $\hat{\Lambda}_T$, namely $\lambda_\pm$, of the $k$-th run. Being the measurement projective, one will collect $N_+$ ($N_-$) outcomes for $\lambda_+$ ($\lambda_-$) such that $N = N_+ + N_-$. The determination of such an expectation value, however, requires the ability to determine $|\nu|$ and $\phi$ together with their derivatives with respect to $T$ with the aim of choosing $\theta=\phi+\varphi$. By tracking $\nu(t)$ one can fix the ``optimal" probing time at which $\mathcal{Q}$ is maximal, and therefore quantify what is the actual uncertainty on the estimated gas temperature. Nonetheless, to determine these quantities a prior knowledge of the temperature of the gas is needed. This means that, before the actual temperature estimation begins, one needs first to determine the probability of obtaining $\xi_k$ in an experimental run, that is, $p(\xi_k\vert T)$, which is no less that the conditioned probability on the actual value, yet unknown, of the gas temperature to be determined. How to estimate the probability $p(\lambda_\pm\vert T)$? This is accomplished by means of the expectation value of $\hat{\Lambda}_T$ (or the energy) as~\cite{Wasak2016}
\begin{equation}
    p(\lambda_\pm\vert T) = \frac{1 \pm \langle \hat{\Lambda}_T \rangle}{2} = \frac{1 \pm f(\langle \hat{E}_T \rangle)}{2},
    \label{eq:pcond}
\end{equation}
where 
\begin{equation}
    f(\langle \hat{E}_T \rangle) = \frac{2\langle \hat{E}_T \rangle - E_{\mathrm{max}} - E_{\mathrm{min}}}{E_{\mathrm{max}} - E_{\mathrm{min}}}.
    \label{eq:fE}
\end{equation}
Here, we used the fact that the expectation value of $\hat{\Lambda}_T$ has support in the interval $[-1, 1]$, while the energy expectation value has been normalized accordingly.  Of course, one needs first to perform a calibration of the ionic thermometer. This can be accomplished either with a theoretical model of the probabilities, as the one based on $\langle \hat{\Lambda}_T \rangle$ given by Eq.~(\ref{eq:SLD}), or by measurements of the energy spectrum as $p(\lambda_\pm\vert T) = N_\pm /N$ for some known values of $T$ and $N\gg 1$. The latter can be extracted independently by looking at specific properties of the gas (e.g., wings of the spatial distribution of the gas in the time-of-flight measurement). After this calibration stage, 
the inference about the value of the unknown gas temperature is drawn from the data $\{\xi_k\}_{k=1}^N$ by means of a certain function $\mathcal{T}$ of the acquired measurement data, which in the context of the theory of parameter estimation it is called estimator. A commonly used one is the maximum likelihood estimator, which is defined as the value of $\mathcal{T}$ that maximizes the joint probability distribution $L(T) = \prod_k p(\xi_k\vert T)$ with respect to $T$. The sequence of $N$ experimental runs is performed several times and yields an estimate of the unknown gas temperature $T$, namely it is expected that its statistical mean, $\mathsf{E}[\mathcal{T}] := T_{\mathrm{est}}$, is $T_{\mathrm{est}} \simeq T$. Since the outcomes of the measurements fluctuate from a data set to another one, i.e., for fixed $N$ the variables $N_\pm$ are stochastic as well as the likelihood $L(T)$, the estimator $\mathcal{T}$ of the unknown gas temperature has an uncertainty.
Given the fact that the protocol gives access to the measurements set by the SLD that maximizes the Fisher information, the uncertainty is then given by the Cramer-Rao bound, that is, the right-hand-side of Eq.~\eqref{eq:SNR_QSNR}. For a more precise mathematical formalism of the estimation of an unknown parameter, we refer to Ref.~\cite{Wasak2018}. Albeit it concerns the estimation of the gradient of a magnetic field, the formalism applies to our context in the precise same way. \\
Finally, let us note that the decoherence function $\nu(t)$ can be determined by means of many-body Ramsey interferometry~\cite{Demler_PRX12}, which has been successfully implemented for the experimental observation of Fermi~\cite{Cetina_SC16} and Bose~\cite{jan2016,cornell2016} polarons. 

\textit{Few-body processes. $-$} The times required to attain the highest values of $\mathcal{Q}$ could lead in experiments to undesired chemical reactions, which are not only not considered in the theory presented in this work, but, importantly, they would affect the state of the bath and thus resulting in a bad temperature sensor. In particular, three-body recombination is the main process owed to the presence of deeply bound states. Estimation of the decay rate $\gamma$ based on classical trajectory theory~\cite{greene2016,jesus2018,jesus2021} predicts a rate on the order of a few Hz for gas densities on the order of 10$^{12}\,$cm$^{-3}$ up to a few kHz for 10$^{14}\,$cm$^{-3}$, therefore from seconds to sub-millisecond timescales. The numbers quoted in table~\ref{tab:values} for the Fermi time and the QSNR of Fig.~\ref{fig:QSNRs} show that we are reasonable good within the predicted decays. We note, however, that since the ion interacts with a spin-polarized fermionic bath, the quantum statistic of the gas helps in this regard, as no more than one atom can populate a two-body bound state~\footnote{We note, however, that when a fermion of the bath is brought to populate a two-body bound state via a three-body collision, there is a finite probability that the released energy may lead to a spin-flip, i.e., a spin impurity in the bath might be created.}. Furthermore, given the fact that very deep bound states are less likely populated, one can conclude that the fermionic statistic of the gas does not allow to populate more than the most loosely bound state of the polarization potential, thus losing eventually a single atom of the bath per ion. This would be not the case if the bath would be bosonic for which mesoscopic molecular ions can be formed~\cite{Coteprl2002,schurer2017,Gregory2021,georg2021}.\\ 
Another important collisional process to be taken into account is spin relaxation. As it has been shown in the experiment of Ref.~\cite{ratschbacher2013}, the ion spin can decohere rather quickly because of spin-exchange and spin-nonconserving interactions. In particular, the spin-orbit coupling provides a major role in the relaxation dynamics of the ion internal state~\cite{Tscherbul2016}. Specifically, it has been observed in a $^{87}$Rb bath that after a few Langevin collision times $t_L$ the probability of finding a Yb$^+$ in the initial spin configuration is close to 15\%~\footnote{The Langevin time is defined as $t_L = 1/\gamma_L$ with $\gamma_L = 2\pi n_g \sqrt{C_4/\mu_{\mathrm{red}}}$ being the energy-independent rate with $n_g$ the gas density.}. For a $^6$Li - $^{174}$Yb$^+$ compound system with a gas density of $6.2\times 10^{12}\,\text{cm}^{-3}$ we find $t_L/\tau_{\mathrm{F}}\simeq 13$ for $k_{\mathrm{F}} R^\star = - 0.5$. This can be indeed a major obstacle to the successful realization of the ion thermometer. A solution is discussed in the next paragraph. In addition to this, we mention that for low-dimensional atom-ion systems such as quasi-one dimensional the impact of spin relaxation can be limited, as the spin-orbit coupling is reduced, while confinement-induced resonances can provide a tool to control atom-ion interactions~\cite{Melezhik2016}.
\\
An alternative strategy that can be employed to avoid undesired few-body processes is given by Rydberg-dressing~\cite{Negretti_PRA16,Gerritsma_PRL17}, where the atomic cloud is slightly coupled to a Rydberg-state. In such a way the atomic polarizability is enhanced and the ion can be placed to some distance (on $\mu$m scale) from the bath, thus without the need to immerse the ion in the latter~\cite{Negretti_PRA16}. This strategy is not only suitable to reduce the impact of micromotion when ions are confined in a Paul trap, but also to reduce the aforementioned spin relaxation effect. If we want to keep the $in$ $situ$ character of the protocol, however, we can still immerse the ion in the Rydberg-dressed atomic bath, but by means of a properly laser-engineered potential of the form~\cite{Gerritsma_PRL17}
\begin{equation}
    V_d(\mathbf{r})=\frac{A\,R_w^4}{r^4+R_w^4}-\frac{C_4}{r^4}.
    \label{eq:V_dressed}
\end{equation}
Here, $A$ and $R_w$ are laser-controlled parameters. With such an engineered atom-ion interaction it is possible to create a repulsive barrier [the first term in Eq.~(\ref{eq:V_dressed})] around the ion to avoid the atoms to get too close to it. Note that recent experiments have shown that long-range and laser-controlled interactions can be realized~\cite{Ewald_PRL19,denschlag2020,haze2019stark} and an ion-induced Rydberg excitation blockade can be realized~\cite{pfau2018}. The  two-photon scheme utilized in the latter can be exploited also to initialize the thermometer, i.e., to create the ion impurity in the atomic cloud in the internal state $\ket{0}$.  Finally, we note that the aforementioned Rydberg-dressing strategy modifies the short-range interaction in the scale of a few tens of nanometers without affecting the potential at micrometer distances. For this reason, it can also help to suppress the eventuality of charge-exchange collisions that may occur at long times~\cite{Saito_PRA17}. 




\section{Conclusions}
\label{sec:Conclusions}

Based on an interferometric protocol for the $in$ $situ$ estimation of the temperature of a Fermi gas with immersed neutral atomic impurities~\cite{Goold_PRL20}, we investigated the performance of the scheme when ions are utilized as thermometers. We have found that the long-ranged character of the atom-ion polarization potential substantially modifies the quantum signal-to-noise ratio and that it enhances the sensor performance, especially when a single interacting internal state is used. We have investigated mainly two scenarios, namely a static ion and an ion ground-state cooled in a finite trap, for various impurity-gas interactions and different number of bound states of the two-body atom-ion potential. We compared our findings with the case of a static neutral impurity, as originally proposed in Ref.~\cite{Goold_PRL20}, whose impurity-bath interaction is described by a zero-range pseudopotential. 
In Sec.~\ref{sec:Thermometry} we provided an alternative scheme that enables to use a single internal state, and therefore to reach higher values of $\mathcal{Q}$ in shorter times ($< 10^3\,\tau_F$). The latter point is important to limit the impact of ion spin relaxation. 
In Sec.~\ref{sec:Results_0BS} we studied the temperature and time dependence of the QSNR for $k_\mathrm{F}a=-0.5$ [see Fig.~\ref{fig:QSNR2D_NBS}(a)] and $k_\mathrm{F}a=-1.5$ [see Fig.~\ref{fig:QSNR2D_NBS}(b)] finding that in order to attain a higher sensor sensitivity the probing time has to be larger compared to the case of a neutral impurity [see Fig.~\ref{fig:QSNR2D_NBS}(c) and~\ref{fig:QSNR2D_NBS}(d)]. We have shown in Fig.~\ref{fig:Qmax}(a) that the dependence of the maximum of the quantum signal-to-noise ratio $\mathcal{Q}$ on $k_\mathrm{F}a$ (i.e., on the gas density) at fixed $T$ is strongly modified by the long-range potential and it presents a peak. The effect of this finding, however, is reduced when both internal states are used for sensing the gas temperature [see Fig.~\ref{fig:Qmax}(b)], since the accumulated phases of the states compensate each other partially. In Sec.~\ref{sec:Results_LowT} we analyzed the impact of a Gaussian spatial density of the ion at low gas temperatures, i.e. $T=0.05\,T_\mathrm{F}$ and $k_\mathrm{F}a=-0.5$, and found that a finite width of the ion trap can enhance the sensitivity of the thermometer as well as reduces the probing time with respect to the case of a delta-shaped spatial density, i.e., static ion approximation (see also Fig.~\ref{fig:QSNRs}). Finally, in Sec.~\ref{sec:Discussion} we discussed the implementation of the interferometric protocol based on the most recent experimental observations involving hybrid atom-ion systems.

In the present study we have focused our attention on a single impurity. It is well-known in quantum parameter estimation theory that entanglement can further enhance the sensitivity of quantum sensors. Thus, in the future it would be interesting to investigate how the estimation bound can be improved by entangling two ions, a task that is routinely accomplished in trapped ion experiments, and to devise novel interferometric protocols to attain the bound in this case. Furthermore, it would be also interesting to explore the impact of the impurity motion in more detail, for instance, by studying the properties of the Green's functions of the system, as recently undertaken in Ref.~\cite{Bruun_arXiv21} as well as via a master equation approach~\cite{Oghittu_PRA21}. Using motional states of the ion for sensing the gas temperature is preferable in view of reducing the impact of few-body processes. 


\section*{Acknowledgements}
This work is supported by the project NE 1711/3-1 of the Deutsche Forschungsgemeinschaft. We are very grateful to T. Wasak for his constructive feedback on our analyses as well as for the insight in the estimation procedure. We also thank R. Gerritsma for feedback on the manuscript. 


\appendix

\section{Dynamics of the ion internal state}
\label{sec:Calculations}
In Sec.~\ref{sec:System} we stated that the Bloch vector is given by $\mathbf{v}=(\mathrm{Re}[\nu],\mathrm{Im}[\nu],0)$. To show this, we consider the time evolution of the density matrix of the composite atom-gas system $\hat{\rho}$. This is given by $\hat{\rho}(t) = \hat{\mathcal{U}}(t)\hat{\rho}(0)\,\hat{\mathcal{U}}^\dag(t)$, where $\hat{\mathcal{U}}(t)=e^{-i\hat{H}t/\hbar}$ with $\hat{H}=\hat{H}_B+\hat{H}_P+\hat{H}_I$ the total Hamiltonian. Specifically, we have
\begin{equation}
    \begin{split}
        &\hat{\rho}(t) =\frac{1}{2}\times\\
        &\!\!\!\begin{pmatrix}
            e^{\frac{i}{\hbar} (\hat{H}_\mathrm{B}+\hat{H}_0) t}\hat{\rho}_\mathrm{B} e^{-\frac{i}{\hbar}  (\hat{H}_\mathrm{B}+\hat{H}_0) t} & e^{\frac{i}{\hbar}(\hat{H}_\mathrm{B}+\hat{H}_1) t}\hat{\rho}_\mathrm{B} e^{-\frac{i}{\hbar}(\hat{H}_\mathrm{B}+\hat{H}_0) t} \\
            e^{\frac{i}{\hbar}(\hat{H}_\mathrm{B}+\hat{H}_0) t}\hat{\rho}_\mathrm{B} e^{-\frac{i}{\hbar} (\hat{H}_\mathrm{B}+\hat{H}_1) t} & e^{\frac{i}{\hbar} (\hat{H}_\mathrm{B}+\hat{H}_1) t}\hat{\rho}_\mathrm{B} e^{-\frac{i}{\hbar} (\hat{H}_\mathrm{B}+\hat{H}_1) t}
            \end{pmatrix}.
    \end{split}
    \label{eq:rhot}
\end{equation}
The components of the Bloch vector are given by the Hilbert-Schmidt inner product $v_i = \mathrm{Tr}[\hat{\rho}(t)\hat{\sigma}_i]$ with $\hat{\sigma}_i$ $i = x, y, z$ being the Pauli matrices, which yield exactly Eq.~(\ref{eq:blochvec}). Let us note that the trace is taken over both the ion and gas degrees of freedom and where we have exploited the cyclic property of the trace.\\
We can then use Eq.~\eqref{eq:rhot} to show that the projective measure of the energy yields the expectation value of the SLD. To this end,
we define the matrix of a $\pi/2$-pulse as
\begin{equation}
    \hat{R}_{\pi/2}(\theta) = \frac{1}{\sqrt{2}}\begin{pmatrix}
        e^{i\theta} & 1 \\
        1 & -e^{-i\theta}
        \end{pmatrix}
\end{equation}
and we calculate $\hat{R}_{\pi/2}(\theta)\hat{\rho}(t)\hat{R}_{\pi/2}^\dag(\theta)$. This gives the following matrix:
\begin{equation}
    \begin{split}
        &\hat{\rho}_\theta(t)=\frac{1}{4}\times\\
        &\begin{pmatrix}
        2+e^{i\theta}\nu(t)+e^{-i\theta}\nu^*(t) & \nu^*(t)-e^{2i\theta}\nu(t)\\[1ex]
        \nu(t)-e^{-2i\theta}\nu^*(t) & 2-e^{i\theta}\nu(t)-e^{-i\theta}\nu^*(t)
        \end{pmatrix}.
    \end{split}
\end{equation}
The projective measure of the energy, that is, on the $\hat{\sigma}_z$ basis, reads
\begin{equation}
    \mathrm{Tr}\big[ \hat{\rho}_\theta \hat{\sigma}_z \big]=\mathrm{Re}[\nu(t)]\cos(\theta)-\mathrm{Im}[\nu(t)]\sin(\theta).
\end{equation}
By choosing $\theta=\phi+\varphi$ we obtain the desired expectation value of the symmetric logarithmic derivative.
 
\section{Remarks on the numerical simulations}
In this section we provide some pieces of information on the employed numerical methods utilized to assess the quantum signal-to-noise ratio, especially with regard to the diagonalization of the impurity-bath Hamiltonian and decoherence function via the Levitov formula.

\subsection{Finite size system}
Let us recall the single-particle Hamiltonians of Eq.~\eqref{eq:h1}:
\begin{equation}
        \hat{h}_\mathrm{B} = -\frac{\hbar^2}{2m}\nabla^2+V_\mathrm{ext}(\mathbf{r}), \qquad
        \hat{h}_\alpha = \hat{h}_\mathrm{B}+V^{(\alpha)}_\mathrm{eff}(\mathbf{r}).
\end{equation}
Despite treating the gas as homogeneous, we solve the eigenvalues equations for a system confined in a sphere with finite size $R$. The value of the latter can be defined by fixing both the density and the number $N_f$ of fermions in the $s$-wave state at $T=0$ giving $N_f/R=\sqrt{2mE_\mathrm{F}}/\pi\hbar$. The number of fermions is chosen large enough to achieve the thermodynamic limit for the considered timescales. Specifically, $N_f=400$ was sufficient.\\
The eigenstates and eigenenergies satisfying $\hat{h}_\mathrm{B}\ket{\phi_n}=\epsilon_n\ket{\phi_n}$ (without $V_{\mathrm{ext}}$) are given by
\begin{equation}
    \langle\mathbf{r}\vert \phi_n\rangle=\sqrt{\frac{1}{2\pi R}}\frac{\sin(k_nr)}{r}, \quad \epsilon_n=\frac{\hbar^2k_n^2}{2m}
\end{equation}
with $k_nR=n\pi$.\\
As far as the eigenstates and eigenenergies of $\hat{h}_1$ are concerned, analytical solutions are available only in the case of the regularized pseudopotential (see the supplemental material of Ref.~\cite{Goold_PRL20} for details). In the case of the ionic impurity with the regularized two-body potential~\eqref{eq:potential}, however, we need to numerically diagonalize $\hat{h}_1$. Towards this end, we used the~\textit{Mathematica} built-in function~\texttt{NDEigensolve}, as it has been proven to be an efficient and versatile tool for such a task.

\subsection{Computation of the decoherence function}
In order to compute the decoherence function $\nu(t)$ we need to evaluate the determinant in Eq.~\eqref{eq:simplified_nu} at each time. Towards this aim, we represent the operator $\hat{\mathcal{M}} \equiv 1-\hat{n}+\hat{n}e^{i\hat{h}_0t/\hbar}e^{-i\hat{h}_1t/\hbar}$, of which the determinant has to be assessed, in the basis of eigenstates $\ket{\phi_j}$ of $\hat{h}_\mathrm{B}$. Hence, the corresponding matrix elements read
\begin{equation}
    \begin{split}
        &\bra{\psi_m}\hat{\mathcal{M}}\ket{\psi_n}=(1- n_n)\delta_{n,m}+\\
        &n_m\sum_l^{\mathcal{N}_0}\sum_k^{\mathcal{N}_1}e^{i(E_l-E_k^\prime)t/\hbar}\bra{\phi_m}\ket{\psi_l}\bra{\psi_l}\ket{\psi^\prime_k}\bra{\psi'_k}\ket{\phi_n}
    \end{split}
    \label{eq:matrix_element}
\end{equation}
where $\ket{\psi_l}$ ($\ket{\psi'_l}$) are the eigenstates of $\hat{h}_0$ ($\hat{h}_1$) with eigenenergy $E_l$ ($E^\prime_l$), and the fermionic occupation number of the $j$-th eigenstate is given by
\begin{equation}
    n_j=\Bigg\{\exp\bigg[\frac{T_\mathrm{F}}{T}\frac{1}{\Bar{\epsilon}_F}\Big(\Bar{\epsilon}_n-\bar{\mu}\Big)\bigg]+1\Bigg\}^{-1}.
    \label{eq:rescaled_n}
\end{equation}
Here, for the sake of numerical convenience, we have rescaled (indicated by an overbar) the energies with respect to $\hbar^2/(2ma^2)$ and lengths with respect to the $s$-wave impurity-bath scattering length $a$. Moreover, $\Bar{\mu}$ denotes the rescaled chemical potential of the Fermi gas that has been determined by solving $\mathrm{Tr}[\hat{n}] = N_f$. In order to determine the right dimension of the Hilbert space $\mathcal{N}_\mathrm{B}$ of $\hat{h}_\mathrm{B}$ such that the desired numerical accuracy has been reached, we proceeded as follows: once the number of fermions $N_f$ has been fixed, we imposed that $\vert\mathrm{Tr}[\hat{n}]-N_f \vert<\epsilon$ with $\epsilon\in (0,1]$. Specifically, we have chosen $\epsilon = 10^{-4}$, which results in a good tradeoff between accuracy and computational time. The dimensions of the Hilbert spaces $\mathcal{N}_{0,1}$ of $\hat{h}_{0,1}$ are varied arbitrarily up to the value at which the result is convergent. We get for them a number of the same order of $\mathcal{N}_\mathrm{B}$. Albeit the value of $\mathcal{N}_\mathrm{B}$ and $\mathcal{N}_{0,1}$ depend on the gas temperature and number of fermions, that is, the higher the temperature, the large is the Hilbert space dimension, all of them range typically between 500 and 1300.


\bibliography{refs}

\end{document}